\documentclass[preprint,12pt]{elsarticle}

\usepackage[pdfpagelabels=false]{hyperref}	
\hypersetup{colorlinks=true,linkcolor=blue,citecolor=blue,filecolor=blue,urlcolor=blue,}

\usepackage{amssymb}
\usepackage{amsmath}
\journal{Astroparticle Physics}

\begin{document}

\begin{frontmatter}



\title{Enhancing the Cherenkov Telescope Array Observatory high-level performance through an event-type-based analysis} %% Article title


\author[1]{J. Bernete\corref{cor1}}%% Author name 
\ead{juan.bernete@ciemat.es}
\author[1]{S. García-Soto}
\author[1]{T. Hassan}
\author[2,3]{O. Gueta}
\author[2]{M. Linhoff}
\author[4]{A. Sinha}
\author[3]{G. Maier}

\cortext[cor1]{Corresponding author}

\affiliation[1]{organization={Centro de Investigaciones Energéticas Medioambientales y Tecnológicas (CIEMAT)},%Department and Organization
            addressline={Av. Complutense 40}, 
            city={Madrid},
            postcode={28040},
            country={Spain}}
\affiliation[2]{organization={CTAO, Science Data Management Centre (SDMC)},
            addressline={Platanenallee 6},
            city={Zeuthen},
            postcode={15738},
            country={Germany}}
\affiliation[3]{organization={DESY},
            addressline={Platanenallee 6},
            city={Zeuthen},
            postcode={15738},
            country={Germany}}
\affiliation[4]{organization={Dept. of High Energy Physics, Tata Institute of Fundamental Research},
            addressline={Dr. Homi Bhabha road, Colaba},
            city={Mumbai},
            postcode={40005},
            country={India}}

%% Abstract
\begin{abstract}

The analysis traditionally employed by Imaging Atmospheric Cherenkov Telescopes involves optimizing quality cuts to select a sub-sample of high-quality events. These events are used for the scientific interpretation of the data, employing a single set of Instrument Response Functions (IRFs). All selected events are treated equally and assumed to be well represented by these IRFs, while the rest are discarded. An alternative approach, successfully applied in experiments such as \textit{Fermi}-LAT, is an event-type-based analysis. This method divides datasets into subsamples, each containing events of a given expected reconstruction quality. IRFs are computed for each subsample independently, improving the accuracy with which IRFs represent the reconstruction quality of each event. The high-level analysis of these subsamples is performed treating them as independent observations, each with their own set of IRFs, and analyzed jointly.

In this work we present a proof-of-concept implementation of an event-type-based analysis for the future Cherenkov Telescope Array Observatory (CTAO) using simulated data. A neural network (specifically a multi-layer perceptron) is trained to predict the direction reconstruction error of each event, and the simulated dataset is divided into event types based on this predicted variable. We compute IRFs for each event type and compare them with those from the standard analysis (without event types). Finally, we simulate observations using these event-type-wise IRFs and analyze them with high-level analysis tools to test the performance of both approaches. This implementation demonstrates notable improvements: 25\% to 50\% boost in spatial resolving power and $\sim$25\% in sensitivity. This boost in performance will have strong implications in the scientific exploitation of the CTAO data, especially in crowded regions such as the Galactic Plane or searching for spectral signatures like Dark Matter annihilation lines.

\end{abstract}

\end{frontmatter}
%%%%%%%%%%%%%%%%%%%%%%%%%%%%%%%%%%%%%%%%%%%%%%%%%%

%%%%%%%%%%%%%%%%% BODY OF PAPER %%%%%%%%%%%%%%%%%%

\section{Introduction}

The Cherenkov Telescope Array Observatory (CTAO)\footnote{www.ctao.org} is one of the next-generation ground-based observatories for gamma-ray astronomy at very-high energies (VHE) from 20 GeV to 300 TeV. It consists of two arrays of Imaging Atmospheric Cherenkov Telescopes (IACTs), one for each hemisphere, with telescopes of three different sizes: Large-Sized Telescopes (LSTs, 23 m diameter), Medium-Sized Telescopes (MSTs, 11.5 m diameter) and Small-Sized Telescopes (SSTs, 4.3 m diameter). The ``Alpha Configuration'' (name of the first construction phase) of the CTAO will have 4 LSTs and 9 MSTs in the Northern Site (Roque de los Muchachos Observatory, Spain) and 14 MSTs and 37 SSTs in the Southern Site (Paranal Observatory, Chile).\footnote{https://www.ctao.org/news/ctao-releases-layouts-for-alpha-configuration/}

The CTAO will surpass the current generation of IACTs in several aspects. Current estimates show an improvement by a factor of 5-10 in sensitivity and at least 25\% better angular resolution \citep{Zanin:2021tx}. These enhancements will enable the CTAO to address a broad range of scientific objectives, as detailed in \citet{cta_book}.

CTAO performance is estimated from detailed Monte Carlo (MC) simulations and described by a set of Instrument Response Functions (IRFs) \citep{cta_irf_2021}. The main IRF components describing the instrument performance for gamma-ray observations are the effective area, the energy dispersion and the point-spread function. These IRFs are necessary to interpret the data acquired by the observatory, specifically for generating high-level scientific results (e.g., spectra or light curves). The methodology to calculate CTAO's IRFs, as well as their detailed description, has been described in other works (see \citet{prod5,Bernl_hr_2013,HASSAN201776,Acharyya_2019}) and is briefly discussed in section \ref{sec:IRFs}. 

The computation and evaluation of these IRFs have been essential in the early phases of CTAO, guiding key design decisions for individual telescopes as well as the selection of array configurations and observatory sites. At the current stage of the CTAO, IRFs are used by scientific software tools, such as \textit{Gammapy} \citep{gammapy:2023} and \textit{ctools} \citep{ctools_2016}, to simulate the performance of the future observatory over specific science cases. 

The analysis traditionally used by IACTs, after reconstructing the energy and direction of the triggered showers and assigning them a likelihood of being a gamma-ray event, applies a list of quality cuts to select those events that maximize sensitivity. The events with insufficient quality are discarded from later stages of the analysis (for an alternative based on a Bayesian approach see \cite{PhysRevD.bayes}). MC simulations of extensive air showers initiated by gamma rays, protons, and electrons, designed to reproduce the conditions of the analyzed observations, are used to optimize the quality cuts and estimate the IRFs. Once the quality cuts are applied to select the events that maximize sensitivity, a single set of IRFs is assigned, describing the average quality of all events. However, the reconstruction quality of these events is known to vary significantly. For instance, showers detected by a larger number of telescopes are generally better reconstructed, resulting in a more accurate determination of their direction and energy.

The \textit{Fermi}-LAT (Large Area Telescope) Collaboration \citep{Atwood_2009} demonstrated that high-level analysis performance can be significantly improved by dividing events into groups, or event types, based on their expected reconstruction performance \citep{atwood2013pass8}. For example, events can be separated by their expected direction reconstruction quality, and specific IRFs can be produced for each event type. Incorporating the event-type-based analysis yields multiple benefits: it increases the sensitivity and total effective area, reduces background contamination, and significantly improves the angular and energy resolution for a subset of the events. Building on the success of event types in \textit{Fermi}-LAT, we expect that applying a similar approach to the CTAO will significantly enhance its capabilities, with angular and energy resolution improvements of up to 50\% for the best-reconstructed 20\% of events, as reported by \cite{Hassan_2021, Bernete_2023}.
%Incorporating this additional information into the likelihood analysis yields multiple benefits: it reduces background contamination, increases the effective area and sensitivity, and significantly improves the angular and energy resolution for a subset of the events.  

The CTAO is expected to devote approximately 40\% of the available observing time over the first ten years of operation with the full arrays to a list of Key Science Projects (KSP). The CTAO Consortium defined a list of highly motivated set of scientific topics and observations, proposing them as the future CTAO KSPs \citep{cta_book}. One proposed KSP is the Galactic Plane Survey (GPS). Given that the CTAO GPS will survey the entire Galactic plane with a sensitivity 5 to 20 times greater than previous searches by existing IACTs \citep{hessgps}, and considering the high density of VHE accelerators expected in this region, the survey will frequently be affected by source confusion. To mitigate this problem, the CTAO GPS would benefit from enhanced angular resolution. Improved capabilities to identify source morphology and distinguish between extended and point-like sources will also be essential for population studies in the VHE regime, as well as for advancing our understanding of cosmic-ray production and propagation.

The CTAO will also play an important role on the observation of transient sources, following rapid multiwavelenght and multimessenger alerts from other observatories. A good capability of reconstructing the origin position of gamma-rays will also be helpful to detect this kind of sources in shorter observation times.

A compelling science case where angular resolution plays a critical role is the study of the intergalactic magnetic field (IGMF). In the presence of a sufficiently strong IGMF, deflections of electron-positron pairs are expected to produce extended gamma-ray halos around distant blazars \citep{Abdalla_2021}. Detecting these halos would provide valuable constraints on the properties of the IGMF in these regions.

One of the key scientific goals of the CTAO will be to detect Dark Matter (DM). The CTAO will have the capabilities to search for evidence of DM by performing indirect searches of gamma-ray byproducts originated in DM annihilation or decay, targeting regions where DM density is expected to be high, such as the Galactic Center, dwarf spheroidal galaxies or the Large Magellanic Cloud \citep{Acharyya_2021, mnras_dwarf, mnras_mag}. Several DM models expect specific spectral features, such as monochromatic lines, box signals or sharp cutoffs. A clear detection of any of these features would serve as evidence of DM if detected \citep{DM_2024}. The enhancement of the sensitivity and energy resolution would boost the CTAO’s discovery prospects for these ‘smoking gun’ DM signals.

In this work we propose an event-type-based analysis for IACTs and evaluate its performance boost when applied to the future CTAO arrays. The methodology followed in this work is described throughout Section \ref{sec:methods}. In Section \ref{sec:results} we evaluate the performance of the proposed analysis for several science cases, and in Section \ref{sec:conclusions} we present our conclusions.

\section{Methods}
\label{sec:methods}

\begin{figure*}
    \centering
    \includegraphics[width=\linewidth]{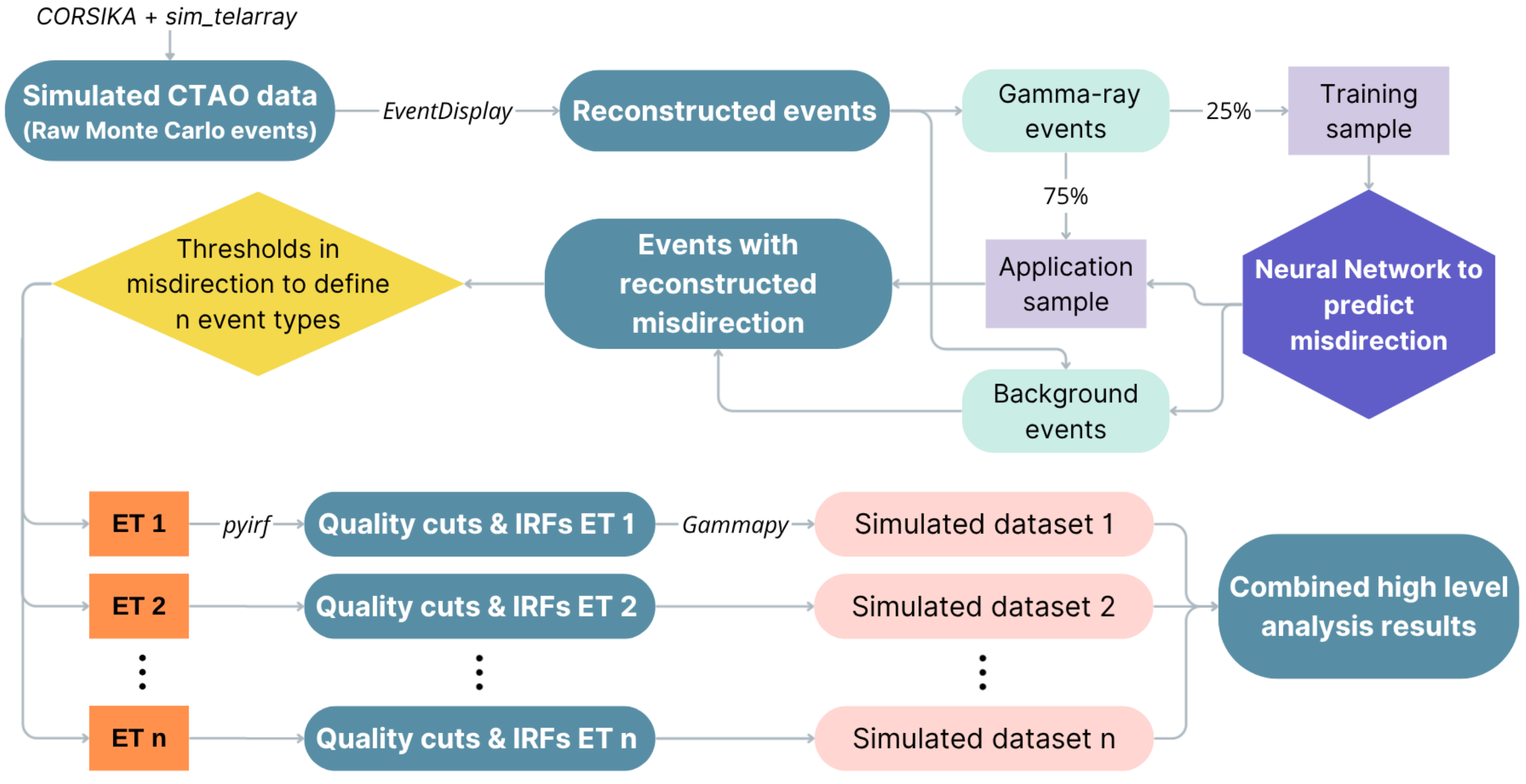}
    \caption{Diagram outlining the steps of the methodology used in this work. ET is used as abbreviation of Event Type.}
    \label{fig:diagram}
\end{figure*}

An outline of the methodology described in the following sections is shown in Figure \ref{fig:diagram}. The simulations and the analysis up to the reconstructed events level are detailed in the following sections (\ref{sec:simulations} and \ref{sec:analysis}). The process carried out to perform the event type separation is described in \ref{sec:et_separation} and the production of the IRFs in \ref{sec:IRFs}. Finally, the high level analysis is presented in Section \ref{sec:high_level}. 

\subsection{Simulations}
\label{sec:simulations}

Monte Carlo simulations of extensive air showers are performed using CORSIKA \citep{corsika_1998} and \textit{sim\_telarray} \citep{simtel_2008} to evaluate the scientific performance of the CTAO. CORSIKA simulates the development of particle showers in the atmosphere initiated by primary particles such as gamma rays, protons, and electrons, while \textit{sim\_telarray} models the response of the CTAO telescopes, including the optics, camera electronics, and trigger systems, under realistic observing conditions.

The simulations used in this study are part of the Prod5 v0.1 dataset \citep{cta_irf_2021, prod5}, which is one of the latest large-scale Monte Carlo productions for the CTAO. These simulations were generated with CORSIKA v7.71, employing QGSjet-II-04 \citep{PhysRevD.83.014018} and URQMD \citep{Bass_1998, M_Bleicher_1999} as hadronic interaction models. The dataset includes gamma rays, protons, and electrons simulated within angular fields of 10 degrees radius around two fixed pointing positions (zenith angles of 20 degrees, pointing directions north and south). The simulations include both the Northern and Southern CTAO arrays, modeled according to the Alpha Configuration \citep{Zanin:2021tx}.

\subsection{Low-level analysis}
\label{sec:analysis}

The analysis from raw simulated data to reconstructed events (generally referred to as ``DL2'', meaning Data Level 2, within the CTAO) is performed using \textit{EventDisplay} \citep{maier2017eventdisplay,2024zndo..11096726M}, identical to the low-level analysis used for generating the CTAO instrument response functions \citep{cta_irf_2021}. It includes calibration, trace integration, image parameterisation, training of boosted decision trees for energy and direction reconstruction, gamma/hadron separation, and stereo event reconstruction. \textit{EventDisplay} parametrizes each stereo image from the simulated events and reconstructs the energy, direction, and \textit{gammaness} (gamma/hadron discrimination parameter with values between 1, meaning closest to gamma-like, and 0, meaning closest to hadron-like).

The analysis proposed in this work would be applicable to any other low-level analysis product with minor modifications (adapting the training features described in the next section to the new data sample).

\subsection{Misdirection prediction and event type separation}
\label{sec:et_separation}

In order to define event types, we need a method to estimate the expected reconstruction quality of each event. In this work, we will use an estimation of the error in the determination of the gamma-ray direction as the parameter representative of their reconstruction quality. This is achieved by predicting the angular difference between the true (simulated) and reconstructed directions, referred to in the following as the misdirection. Misdirection is predicted, and shown along this work, in logarithmic scale (all logarithms in this work are base 10).

To perform this prediction, various machine learning algorithms from the \textit{scikit-learn} library \citep{scikit-learn} were tested, such as random forests, support vector regressors, and multi-layer perceptrons (MLPs). Among these, the MLP with a \texttt{tanh} activation function and two hidden layers (36 and 6 neurons, respectively) provided the best results.  Along with the activation function, we set a maximum of $2\cdot10^4$ iterations and a tolerance of  $10^{-5}$.  The rest of the parameters of the MLP are are set to the default configuration\footnote{http://scikit-learn.org/stable/api/sklearn.neural\_network.html}.

The regression model uses a variety of input features that describe both telescope-level image parameters and array-level ones. A complete list of all parameters used to train the model, along with their definitions, can be found in \ref{app:train_features}. In order to study which variables are most valuable for the regression model, the importance of the parameters is computed using Garson's algorithm \citep{garson_1991, GOH_1995}. The results, available in Figure \ref{fig:importance}, show that some of the most important reconstructed features are the average distance between the centroid of the shower image and the average camera offset, the average distance from each telescope to the reconstructed shower core, the emission height, and the average image size. These high-importance variables reflect the geometrical relationships between the shower images and the array, which strongly affect how well the shower direction is reconstructed.

The misdirection prediction model is trained and tested separately in 20 quantile-based bins of reconstructed energy so each bin contains approximately the same number of events. This approach accounts for the fact that the relationship between input parameters and misdirection varies significantly across energy ranges, improving the performance of the model as it allows it to better capture the energy-dependent patterns in the data. In this work, the 25\% of the simulated data is used for the training and the remaining 75\% constitutes the test sample (see Figure \ref{fig:diagram}). 

An alternative approach to the regression model described above would be to treat the task as a classification problem, labeling events directly into discrete types with no intermediate regression. However, regression is more flexible because it forecasts a continuous quality parameter that can be later partitioned into event types according to different thresholds, thus preserving information about reconstruction accuracy \citep{Hassan_2021}. Furthermore, low-level analysis methods, such as the one described by \citet{schwefer_2024}, could naturally produce such quality parameters, making them directly applicable for event type separation.

\begin{figure*}
    \centering
    \includegraphics[width=\linewidth]{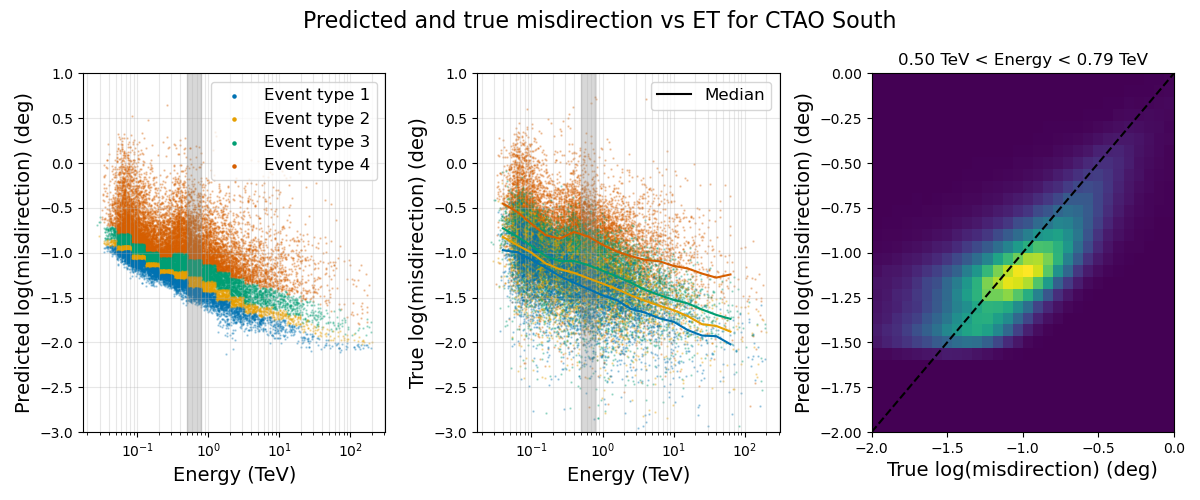}
    \caption{\textit{Left}) Predicted logarithm of the misdirection for gamma-ray events as a function of reconstructed energy. The assigned event type is shown in different colors. \textit{Middle}) True logarithm of the misdirection for gamma-ray events as a function of reconstructed energy. Colors represent the assigned event types. Solid lines indicate the median of the distribution of each event type. \textit{Right}) Predicted vs true logarithm of the misdirection for gamma-ray events with a reconstructed energy between 0.50 and 0.79 TeV, as an example. The gray vertical band in the left and middle plots indicate the energy bin shown in the right plot. All of these plots are limited to a single offset angle bin, from 0 to 1 deg from the center of the FOV.}
    \label{fig:predicted_vs_true_misdirection}
\end{figure*}

Once the misdirection is predicted, events are grouped into event types based on their predicted reconstruction quality. Thresholds are applied to divide the ranked events into consecutive groups, with the best-reconstructed events (smallest predicted misdirection) assigned to ``event type 1'', followed by ``event type 2'', and so on. These thresholds are determined independently in bins of reconstructed energy and angular offset from the center of the field of view (FOV), as shown in Figure \ref{fig:predicted_vs_true_misdirection}. Thresholds are chosen so that each event type contains a fixed fraction of the gamma-ray events; the procedure used to derive the optimal fractions and obtain the final thresholds is described in Section \ref{sec:res_partitioning}. The separation into event types is applied to all gamma-ray, proton, and electron samples. While the gamma-ray proportions in each event type are predefined, the distribution of protons and electrons across event types are expected to vary. The number of event types and the proportion of gamma-ray events in each group can be adjusted depending on the high-level analysis goal or science case.

Figure \ref{fig:predicted_vs_true_misdirection} shows the performance of the misdirection prediction: the left figure shows how event-type thresholds are defined as a function of reconstructed energy, the middle figure shows how the true (simulated) misdirection of the already classified events behaves, while the right figure shows, as an example, the true vs predicted misdirection on a specific reconstructed energy bin (between 0.50 and 0.79 TeV).

\subsection{Production of Instrument Response Functions}
\label{sec:IRFs}

The IRFs provide a mathematical description that links the reconstructed quantities of a list of events to the true physical quantities of the incident photons. The response for the CTAO is factorized into three independent functions: the effective area, the point-spread function and the energy dispersion. An additional function to describe the background rate is also usually included. These components are typically stored in 2D matrices with the axes being energy and FOV offset angle.

The CTAO's IRFs are computed from simulated samples of reconstructed events. The standard methodology to compute them \citep{Bernl_hr_2013, Acharyya_2019, HASSAN201776} starts with a re-weighting of the simulated events so that they resemble the expected distribution from a CTAO observation of a Crab-Nebula-like source (as a test case). Subsequently, a quality cut optimization is needed, generally to get the best sensitivity as a function of the reconstructed energy. Typically in VHE gamma-ray astronomy the sensitivity is defined, for each reconstructed energy bin, as the flux associated to the minimum number of gamma-ray events that meets all of these criteria: the Li\&Ma significance \citep{LiMa_1983} is $5\sigma$ or higher, the number of excess events is 10 or higher and is above $5\%$ of the background counts. The final set of IRFs is computed using the events that survived the quality cuts. The cut optimization is usually performed over the following parameters: multiplicity (number of telescopes used in the reconstruction of an event), \textit{gammaness} and, in the case of a point-like source analysis, the angular size of the signal region (\textit{ON region} with angular radius $\theta$).

Depending on whether $\theta$ cuts are applied or not, the IRFs can be full-enclosure (no $\theta$ cuts applied) or point-like ($\theta$ cuts applied). Full-enclosure IRFs are needed to perform spectro-morphological studies, the so called \textit{3D analysis}, allowing to identify the best-fit model matching the data not only in reconstructed energy, but also in the two direction coordinates. For the point-like IRFs presented in this work, the theta, multiplicity and gammaness cuts are optimized in every reconstructed energy bin by computing the combination of the three parameters that results in the best sensitivity.

In the ``standard'' analysis we just described, all events not surviving the quality cuts are discarded, and not used at all in later stages of the analysis. In addition, by applying a single set of IRFs for all surviving events, one assumed they have identical quality (even if we know it is not true).

In an event-type-based analysis, the partitioning (as explained in Section \ref{sec:et_separation} and summarized in Figure \ref{fig:diagram}) occurs before optimizing the cuts and computing the IRFs. This allows to create a number of independent lists (as many as event types), each with their corresponding set of IRFs describing their average quality. The procedure to produce the event-type-wise IRFs is identical to the one just described, in this case using only a subsample of the simulated dataset, corresponding to one event type, as input for each set of IRFs.

The software used to compute IRFs and store them in GAD format \citep{gamma_data_formats, deil2017open} is \textit{pyirf} \cite{pyirf}, a library specialized in cut optimization, estimation of the sensitivity and generation of IRFs for the CTAO. This library, for which a custom version is used in this work, was first tested and validated to produce offset-dependent and full-enclosure IRFs. These tests are detailed in \citet{Bernete_2023} and resulted in a reasonable agreement between \textit{pyirf} and \textit{EventDisplay}.

\subsection{High level analysis}
\label{sec:high_level}
 
A high-level analysis typically involves binning the events into a counts map and interpolating the IRFs on the chosen analysis geometry. Then physically meaningful information is extracted, such as flux points, light curves, morphologies, etc. (labeled DL5). In this work, we use Gammapy v1.1 \citep{gammapy:2023, gammapy_zenodo_1.1, gammapy_zenodo} to simulate observations (both from the standard IRFs as well as the event-type-wise IRFs we produced), and then use them for benchmarking three science cases: the estimation of sensitivity curves, the detection of the extension of a source and the separation of two close sources.

When dealing with datasets containing event types, the observation associated to each event type is simulated independently. These observations are then combined in a joint analysis. To evaluate the impact of event types, we apply the same analysis with and without event types for different science cases, and compare the achieved performance between them. 

In the following subsections, we discuss three high-level analysis tests which correspond to the science cases introduced above, designed to assess the improvement of both the resolving capabilities and combined sensitivity of the CTAO.

%As we expect the event-type-based analysis will improve both the resolving capabilities and combined sensitivity of the CTAO, the high-level analysis tests we perform are: a combined sensitivity estimation per energy bin, detection of the extension of a source and the separation of two close sources.

\subsubsection{Sensitivity estimation}
\label{sec:hl_sensitivity}

The estimation of sensitivity curves is essential to describe the impact of the event-type methodology as differential sensitivity is often used as the primary descriptor of the potential scientific performance of an array \citep{Bernl_hr_2013}. The standard methodology to compute sensitivity for the CTAO was introduced in section \ref{sec:IRFs}. We will refer to this methodology as the \textit{Li \&  Ma methodology}.

\textit{Gammapy} has a tool (\textit{SensitivityEstimator}) that allows to estimate the sensitivity for a single set of IRFs applying the \textit{Li \&  Ma methodology}. This methodology does not allow to naturally calculate the sensitivity for a joint set of observations with different IRFs. Therefore, another approach is needed to estimate the joint sensitivity of an event-type-based analysis. 

Thus, using \textit{Gammapy}, we create a dataset for each set of point-like IRFs that only contains the expected background, and then we compute 5-$\sigma$ joint flux upper limits (with \textit{Gammapy}'s \textit{FluxPointsEstimator}) for a spectral model mimicking the Crab Nebula (power law spectrum with index=2.62). This method is an approximation of a forward-folding method to estimate joint sensitivity, but it is expected to predict worse performances in particular in regions where the number of counts is low. We will refer to this methodology as the \textit{Forward folding methodology}. Other drawbacks of this method is that it does not allow to set constrains on the minimum number of signal counts or the background fraction. In addition, Cash statistics \citep{1979ApJ_Cash} (default in \textit{Gammapy} for a likelihood ratio test) are used to perform the minimization, which will not be consistent nor directly comparable with Li \& Ma significances. For these reasons, the sensitivity curves computed with this method (especially at the highest and lowest energies) may not be directly comparable to public performance curves. However, since this method is the only one available to use in a joint analysis, this method will be used throughout this work for the calculation of sensitivities using \textit{Gammapy}.

\begin{figure}
    \centering
    \includegraphics[width=\linewidth]{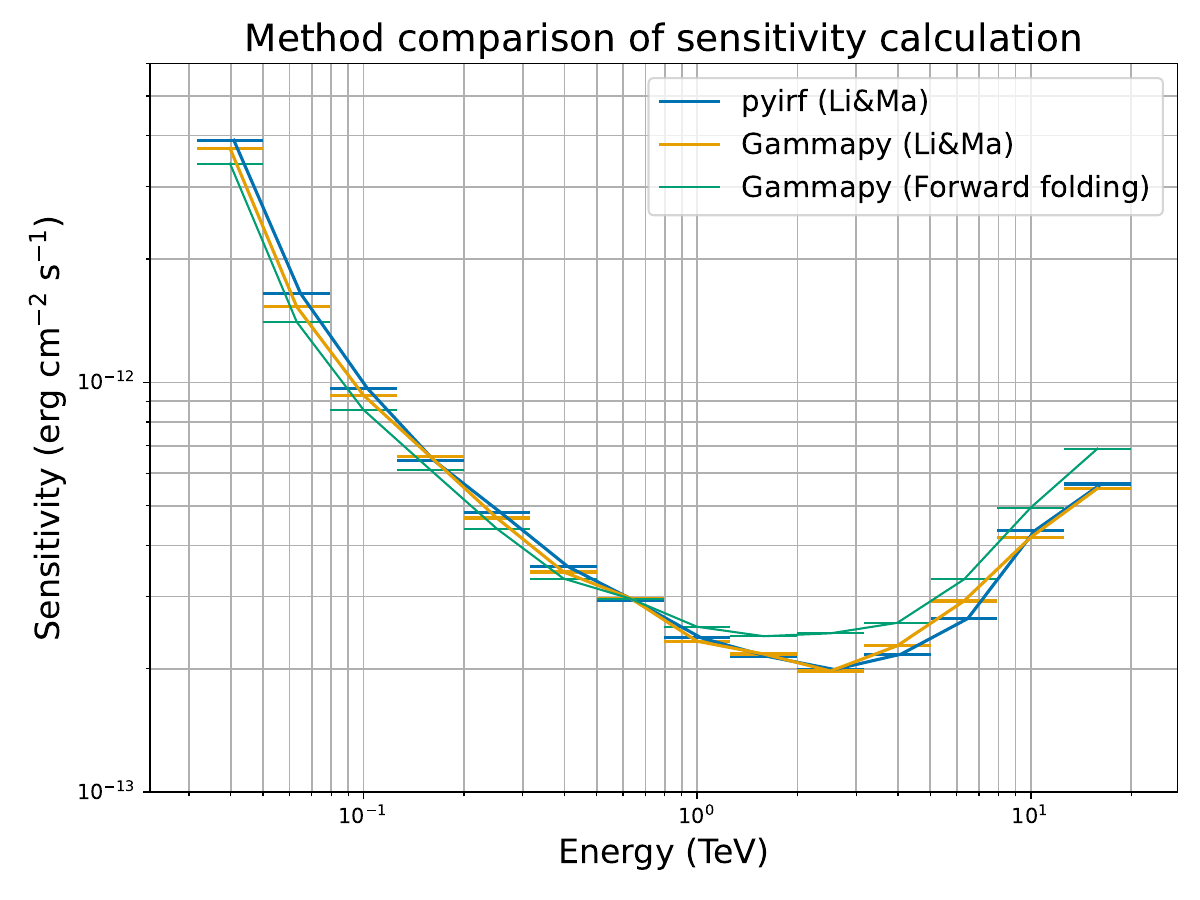}
    \caption{Comparison of the differential sensitivity computed with the 3 different methods described in this work. Sensitivity curves correspond to the CTAO-North array, averaged north and south pointing in 50 h of observing time.}
    \label{fig:sensitivity_methods}
\end{figure}

Figure~\ref{fig:sensitivity_methods} compares the sensitivities computed from the 3 methods described in this work: 1) \textit{pyirf} using the \textit{Li \&  Ma methodology} directly from events, 2) \textit{Gammapy} using an analog \textit{Li \&  Ma methodology} from IRFs and 3) \textit{Gammapy}, using the \textit{Forward folding methodology}. Even when the \textit{Forward folding methodology} has no restriction in the number of signal counts, it shows worse sensitivities for the highest energy bins. This happens because, when zero counts are measured (the case for these bins), the 5$\sigma$ upper limit in Cash statistics is 12.5 counts. The \textit{Li \&  Ma methodology} in this regime is limited by the 10-signal-counts condition, so the \textit{Forward folding methodology} shows fluxes 25\% higher for this high energy bins.

To evaluate the performance of a sensitivity curve in comparison to a reference sensitivity, the performance per unit time (PPUT) is computed \citep{HASSAN201776}. The PPUT is the geometric mean through individual energy bins of the inverse of the normalized sensitivity: 
\begin{equation}
    \text{PPUT}=\left( \prod_{i=1}^{N} \frac{F_{\text{sens,ref}}(i)}{F_{\text{sens}}(i)} \right)^{1/N}\, ,
    \label{pput}
\end{equation}
where N is the number of energy bins, $F_{\text{sens}}$ is the achieved sensitivity and $F_{\text{sens,ref}}$ is the reference sensitivity used for the normalization. In this case, this is calculated for the joint sensitivity of all event types against a reference sensitivity with no event types, both of them obtained with \textit{Gammapy} and evaluated only in the energies where this method is reliable (i.e.: avoiding extreme bins where sensitivity would need to be determined by the conditions on the minimum number of signal counts or the background fraction). We restrict the calculation to an energy range from 30 GeV to 20 TeV for the North site and from 50 GeV to 80 TeV for the South site.

A similar figure can be defined in relation to the angular resolution to compare the performance of several event types configurations. The Angular Performance \citep[AP,][]{HASSAN201776} is calculated in an analogous way to the PPUT as a geometric mean over energy bins of the inverse of the normalized angular resolution: 
\begin{equation}
    \text{AP}=\left( \prod_{i=1}^{N} \frac{\Theta_{\text{0.68,ref}}(i)}{\Theta_{\text{0.68}}(i)} \right)^{1/N}\, ,
    \label{angularperformance}
\end{equation}
where N is the number of energy bins, $\Theta_{\text{0.68}}$ is the computed angular resolution and $\Theta_{\text{0.68,ref}}$ is the one used as reference for the normalization. The angular resolution is defined as the 68\% containment radius and the reference value is obtained from the simulations with no event types applied. 

These quantities are used as figures of merit of the science performance achieved by the event-type-based analysis and are useful to easily compare the different configurations (number of event types and proportion of events in each one). The results obtained for both the PPUT and AP of our analysis are presented in Section \ref{sec:results}.

\subsubsection{Extension detection}
\label{sec:hl_extension}

A common challenge in studying crowded regions, such as the Galactic plane, is in distinguishing overlapping and extended sources. To understand if the use of event-types can aid such studies, we test the performance for the detection of the extension in a source. An observation of an extended source is simulated with \textit{Gammapy} from each set of full-enclosure IRFs (with and without event types). Full-enclosure IRFs are used to conduct a 3D analysis where no on-region needs to be defined. This is also the case for the test presented in the following section (\ref{sec:hl_separation}). The simulated source model follows a Gaussian spatial distribution:
\begin{equation}
    \phi\text{(lon, lat)}=\frac{1}{2\pi\sigma^2}\exp\left(-\frac{1}{2}\frac{\theta^2}{\sigma^2}\right) \, ,
    \label{gaussianmodel}
\end{equation}
where $\theta$ is the angular separation to the model center and $\sigma$ is the effective radius of the Gaussian (equation \ref{gaussianmodel} assumes the limit of both $\theta$ and $\sigma$ being small), and a power law spectral model:
\begin{equation}
    \phi(E)=\phi_0\cdot\left(\frac{E}{E_0}\right)^{-\Gamma }\, ,
    \label{powerlaw}
\end{equation}
with reference energy $E_0=1\text{ TeV}$ and spectral index $\Gamma=2.62$, to resemble the Crab spectral-index as measured by HEGRA \citep{Aharonian_2004}. The flux normalization, $\phi_0$, can be set to different values to test the performance at different flux levels.

Other specifications of the \textit{Gammapy} simulations are: 50 hours of observation time, 50 realizations for each set of parameters, and size of the simulated region set to a square of length 6 times larger than the radius of the source in each case and this region is binned into 40\,$\times$\,40 pixels. The range of simulated source radius goes from 0.006 to 0.3 deg, which correspond to pixel size values from 0.005 to 0.045 deg. Having the size of the simulated region and the pixels scaled with the size of the source allows for a more stable computation and fair comparison. 

We then fit the simulated dataset using: 1) a point spatial model and 2) a Gaussian spatial model, and compare the two using a likelihood ratio test. For all models, the source position, the spectral normalization and the normalization of the FoV background model is kept free. Additionally, for the Gaussian spatial model, the extension is also free. The spectral index is always kept frozen to 2.62 (Crab-like spectrum). 

%For both models, the normalization parameters of the power law spectral model and FoV background model are also free.

\subsubsection{Separation of close sources}
\label{sec:hl_separation}

We simulate two point-like sources in the same field of view with the same spectral model as before, varying the angular distance between them to evaluate the separation capabilities.

All the specifications of these simulations are the same as for the extension detection tests, explained in the previous subsection (\ref{sec:hl_extension}), except that in this case, the size of the simulated region is a square of length 5 times larger than the distance between the two simulated sources.

In this case we fit the simulated dataset to: 1) a one point-like source model and 2) a two point-like sources model with free separation between them, and compare the two using a likelihood ratio test.

As both the extension and source separation tests rely on simulations that introduce randomness into the results, these tests must be repeated a large number of times to reduce statistical fluctuations and obtain a more stable estimate of the mean significance.

\section{Results}
\label{sec:results}

The primary performance criteria for evaluating the event-type-based methodology is the differential sensitivity, calculated using five bins of equal logarithmic width per energy decade and 1 degree wide bins of FOV offset angle. All sensitivity curves and IRFs shown in this section are derived from the same simulated dataset, as described in Section \ref{sec:simulations}, and follow the methodologies outlined in Sections \ref{sec:IRFs} and \ref{sec:hl_sensitivity}. Sensitivity curves for each event type, as well as the combined sensitivity, are presented alongside the IRFs, PPUTs and APs. To assess the impact of the proposed event-type approach to the high-level analysis, two specific tests have been conducted: the detection of source extensions and the angular separation of closely spaced sources.

\subsection{Event-type analysis performance: combined sensitivity}
\label{sec:res_sensitivity}

\subsubsection{Sensitivity}
\begin{figure*}
    \centering
    \includegraphics[width=\linewidth]{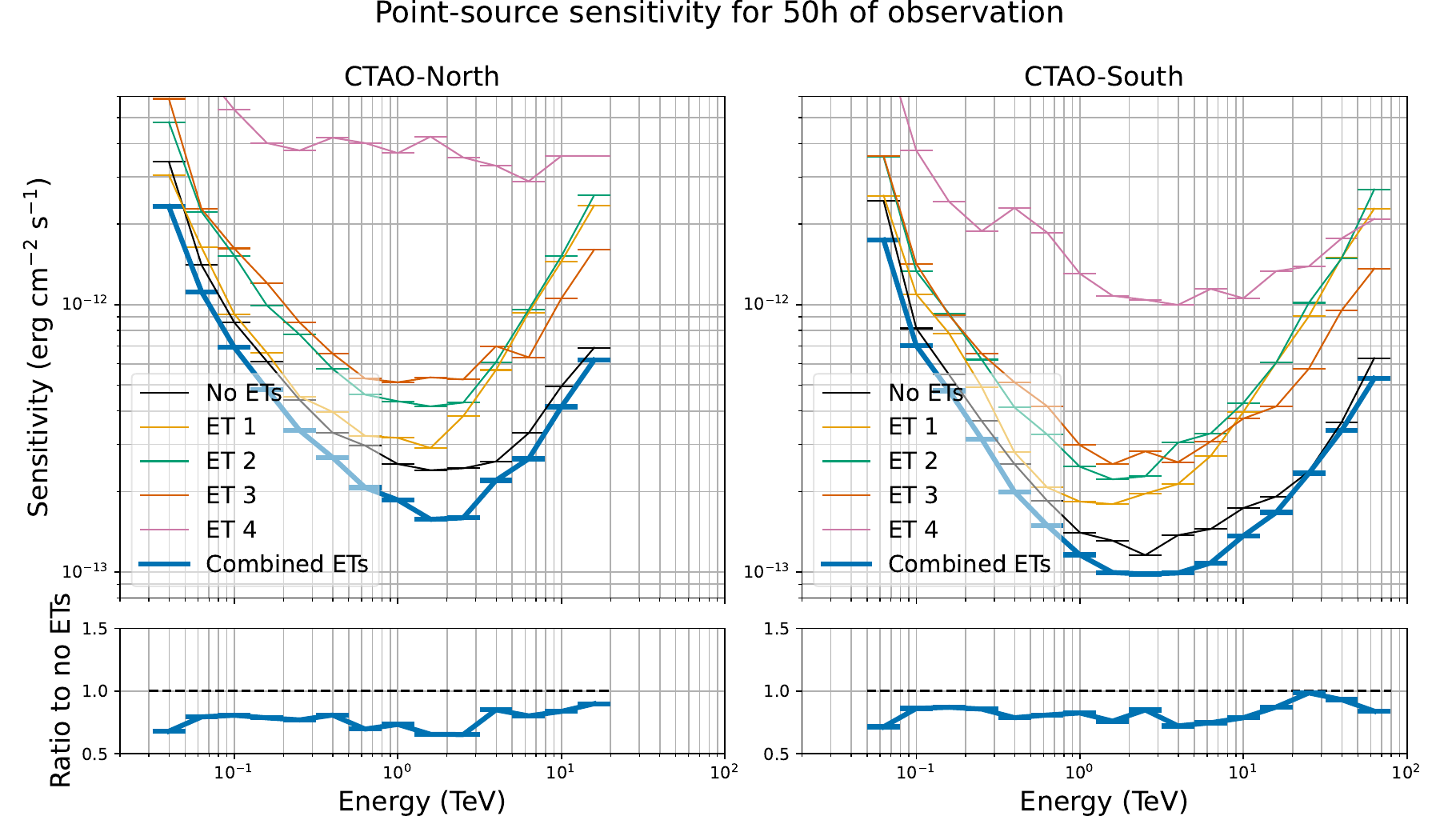}
    \caption{Point-source sensitivity for both arrays at offset 0.5 degrees from the pointing direction. On the top, the sensitivity for each event type separately, for all of them combined and for no event types, for both arrays. On the bottom, the ratio of the combined sensitivity with respect to the sensitivity for no event types. In all panels, lower sensitivity or ratio values indicate better sensitivity. All sensitivities showed here are calculated with the \textit{Forward folding methodology} in \textit{Gammapy}. For both arrays the Alpha Configuration is considered.}
    \label{fig:sensitivity}
\end{figure*}

 The events were divided into four types, with thresholds established to allocate gamma-ray events into proportions of 15\%, 15\%, 30\%, and 40\% respectively, from better to worse reconstruction quality (proportions will be justified later). The sensitivity curves for each type, as well as the combined sensitivity, are shown in Figure \ref{fig:sensitivity}.

The sensitivity for individual event types is worse than the sensitivity without event-type separation, as each type contains only a fraction of the total events. This effect is particularly pronounced at higher energies, where sensitivity is mostly limited by the number of signal events, and event types with fewer events are more affected. At energies below 1~TeV, in the background-dominated regime, the best-quality event type (ET 1) achieves a sensitivity comparable to the standard analysis, while subsequent event types progressively worsen.

The combined sensitivity, calculated by joining all event types, demonstrates a 20-40\% improvement for the Northern array and a 15-30\% improvement for the Southern array over the standard analysis in energies below $\sim$5~TeV. As described in section \ref{sec:hl_sensitivity}, the method used to calculate the combined sensitivity is not directly comparable to public performance curves, but allows a fair comparison between the methods tested. This comparison consistently favors the event-type-based analysis. This indicates that separating events into types based on reconstruction quality allows for a more efficient use of the data, retaining events that would otherwise be discarded in a standard analysis. These results demonstrate that using these otherwise-discarded events allows to improve sensitivity.

\subsubsection{Event-type partitioning optimization}
\label{sec:res_partitioning}
\begin{figure*}
    \centering
    \includegraphics[width=\linewidth]{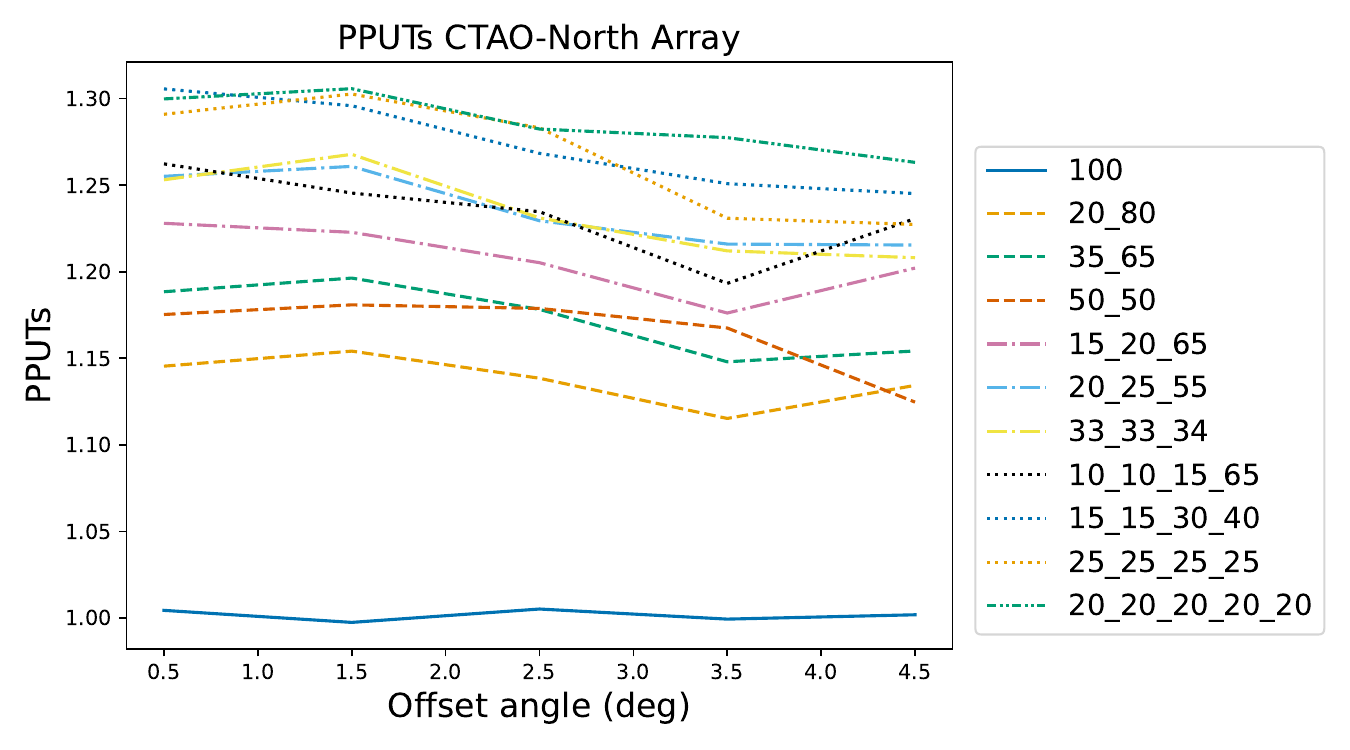}
    \includegraphics[width=\linewidth]{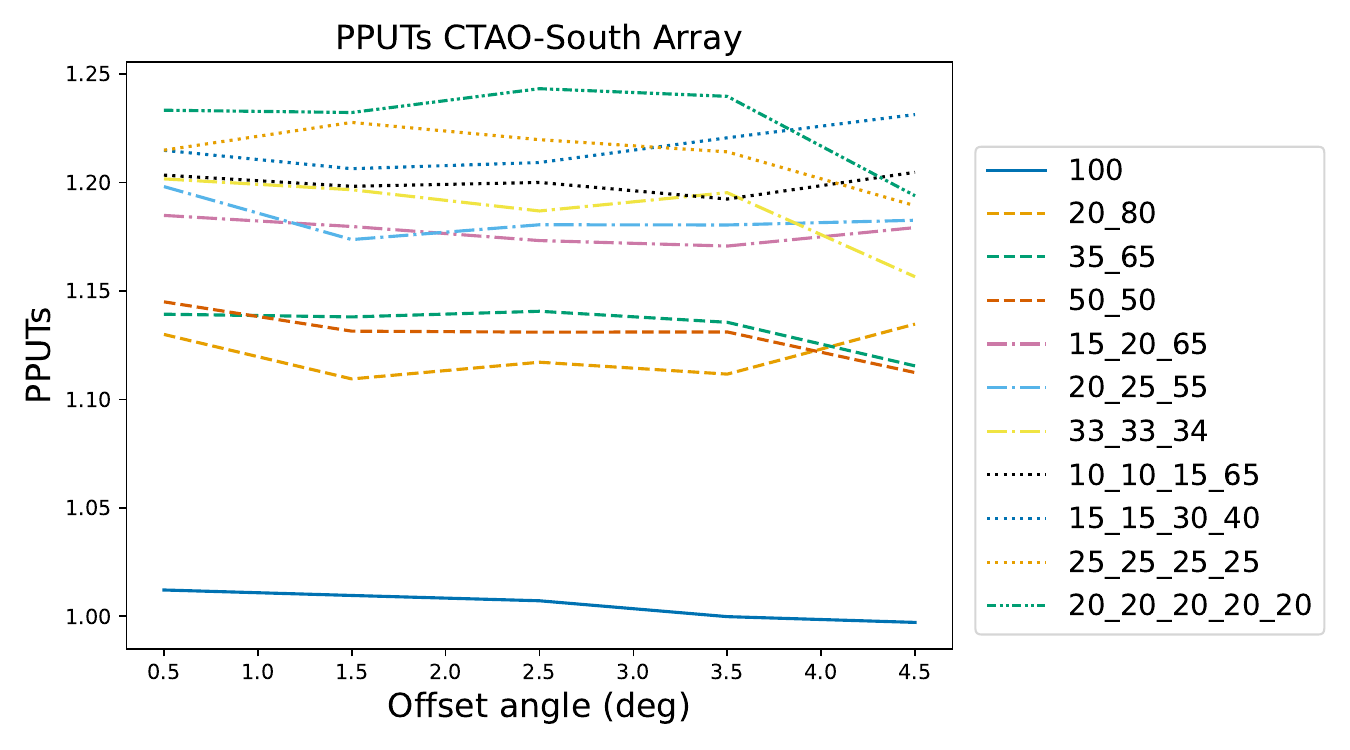}
    \caption{PPUTs for both arrays showing the general improvement of the combined sensitivity of different partitions of event types against the sensitivity of no event types (equivalent to a single event type with 100\% of events), for different offset angles. Labels indicate the proportion of gamma-rays in the different event-type partitions, from better quality to worse. PPUTs from a single event type may deviate from 1 because $F_{\text{sens,ref}}$ was computed with the full Prod-5 statistics, while event-type-wise IRFs are computed with the 75\% of gamma-ray event statistics.}
    \label{fig:PPUTs}
\end{figure*}

The improvement in the combined sensitivity achieved through event-type separation was evaluated for various partitions of gamma-ray event statistics: we explore separations of 2, 3, 4 and 5 event types, as well as several gamma-ray proportions between them. We use the PPUT metric, as defined in Equation \ref{pput}, to compare their performance. Figure \ref{fig:PPUTs} illustrates the PPUT values for different event-type configurations at varying offset angles and Table \ref{table:pputs} shows the mean value of the PPUTs for the same event-type configurations across all the offset values from 0.5 to 4.5 degrees.

\begin{table}
    \begin{tabular}{lll}
    Event type partition & Mean PPUT North & Mean PPUT South \\
    20\_20\_20\_20\_20   & 1.285           & 1.228           \\
    15\_15\_30\_40       & 1.273           & 1.216           \\
    25\_25\_25\_25       & 1.267           & 1.213           \\
    10\_10\_15\_65       & 1.233           & 1.200           \\
    20\_25\_55           & 1.235           & 1.183           \\
    33\_33\_34           & 1.234           & 1.187           \\
    15\_20\_65           & 1.207           & 1.178           \\
    35\_65               & 1.173           & 1.134           \\
    50\_50               & 1.165           & 1.130           \\
    20\_80               & 1.138           & 1.121           \\
    100                  & 1.001           & 1.005          
    \end{tabular}
    \caption{Mean values of the PPUTs for the tested event-type configurations across all the offset values from 0.5 to 4.5 degrees, for both the North and South arrays.}
    \label{table:pputs}
\end{table}

\begin{figure*}
    \centering
    \includegraphics[width=\linewidth]{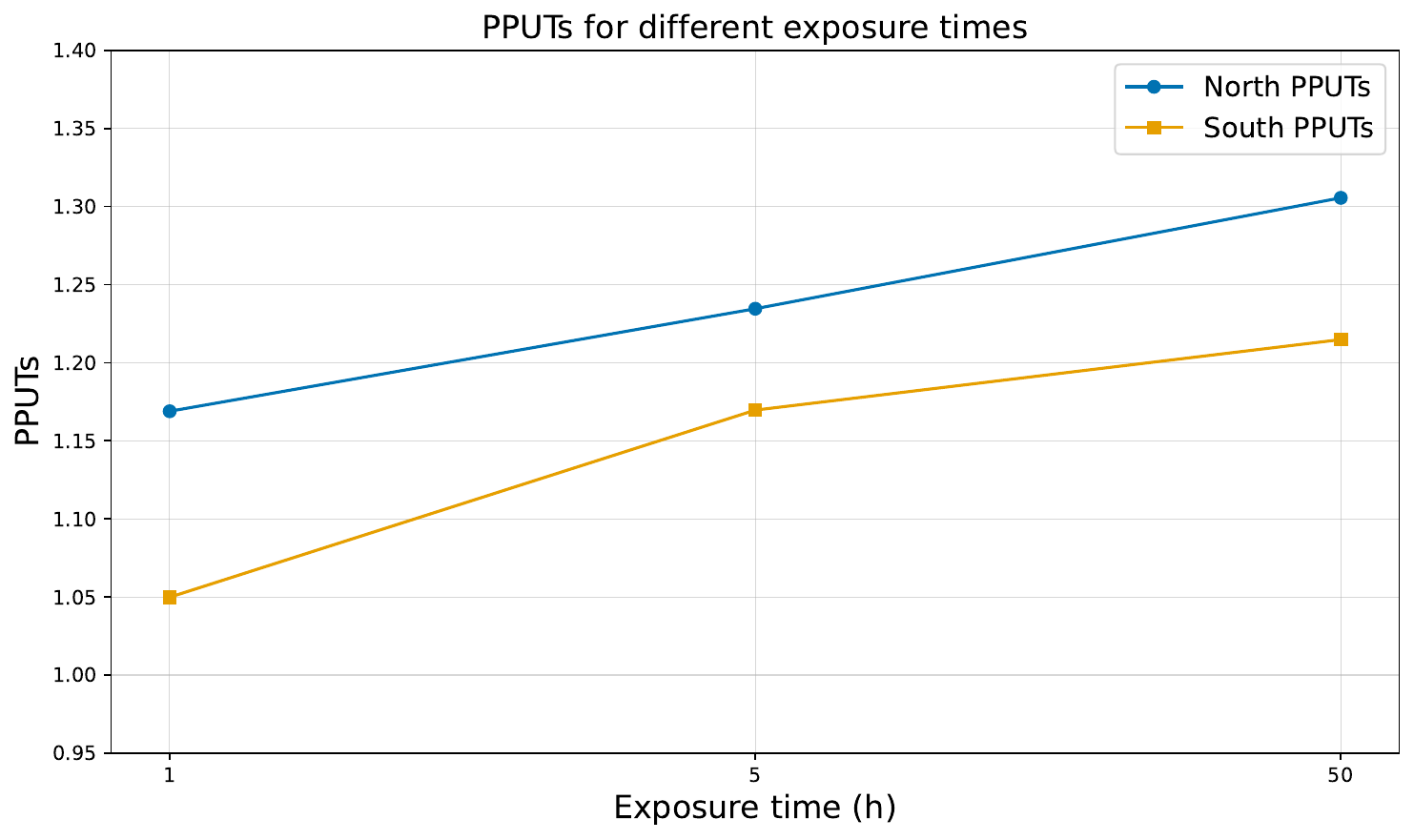}
    \caption{PPUTs for both arrays showing the general improvement of the combined sensitivity of different partitions of event types against the sensitivity of no event types for different exposure times. Only the selected partition of four event types (15\%, 15\%, 30\%, and 40\%) at an offset of 0.5 deg is shown.}
    \label{fig:PPUTs_exposure}
\end{figure*}

The analysis shows that sensitivity improvements occur consistently across the whole FOV. A trend is observed where a higher number of event-type partitions lead to better performance. However, increasing the number of event types beyond four or five may lead to a significant increase in statistical uncertainties due to the finite size of the MC sample. For this reason, the partition with four event types that shows overall better PPUT values (15\%, 15\%, 30\%, and 40\%) was selected, as it balances performance improvements with practical considerations. This configuration achieves a sensitivity enhancement of approximately 25\% compared to the standard analysis. For shorter exposure times, the improvement is lower, but still significant, as shown in Figure \ref{fig:PPUTs_exposure}.

As an additional validation, the data were randomly divided into event types without any input from the misdirection prediction model. The resulting combined sensitivity was found to be consistent to the standard analysis, confirming that the sensitivity improvement is indeed due to the additional information provided by the event-type classification based on the predicted misdirection.

\subsection{Event-type analysis performance: IRFs}
\label{sec:res_IRFs}

Section \ref{sec:IRFs} provides a detailed explanation of the methodology used to produce IRFs within the framework of an event-type-based analysis. We use this methodology to calculate the effective area, angular and energy resolution and background rate for the selected partition of event types (15\%, 15\%, 30\% and 40\%) and their combination. They are all obtained for 50h of observations both for the CTAO North and South arrays.   % we tested improvement stays also at shorter observations times etc etc %% añadir figura similar a la figura 5

\begin{figure*}
    \centering
    \includegraphics[width=\linewidth]{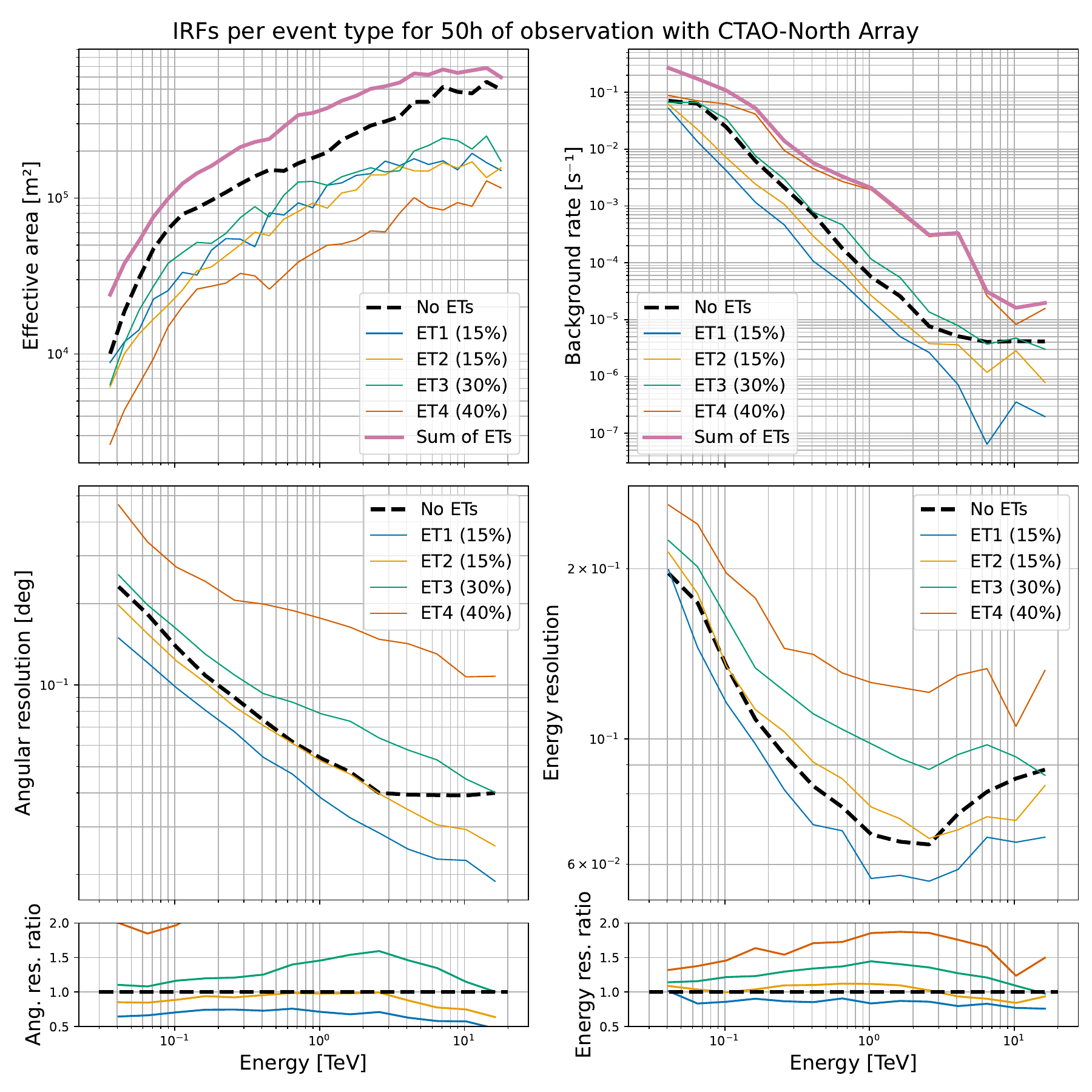}
    \caption{Effective area, background rate, angular resolution and energy resolution for each event type separately and for the standard case without event types, for the CTAO Northern array. They all correspond to point-like IRFs, in which a direction cut optimizing point-source sensitivity has been applied. Effective area and background rate also show the sum of the four event types. Angular and energy resolution also show the ratio of each event type result with respect to the standard case. All of these are for offset angle between 0 and 1 degree.}
    \label{fig:IRFs_n}
\end{figure*}

\begin{figure*}
    \centering
    \includegraphics[width=\linewidth]{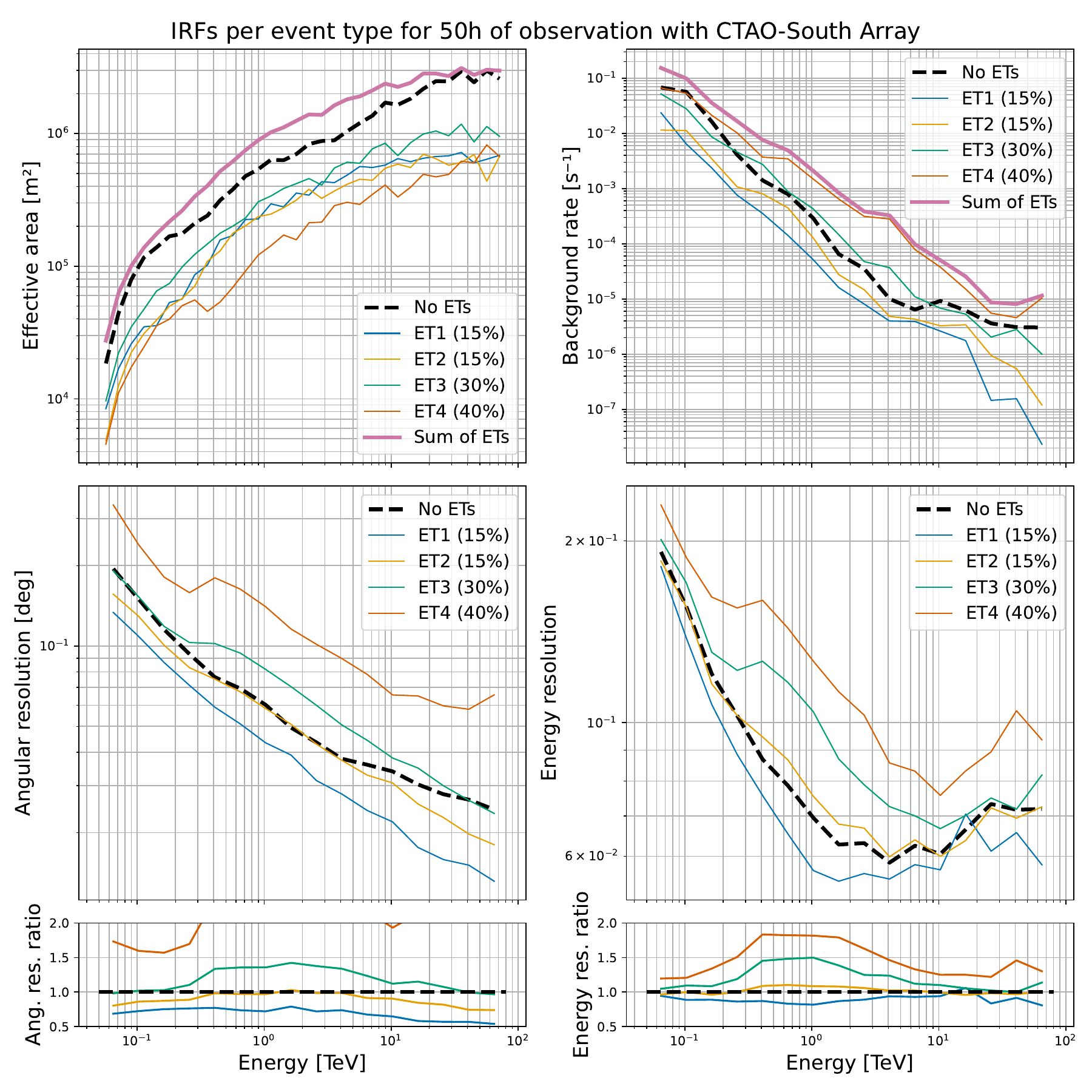}
    \caption{Effective area, background rate, angular resolution and energy resolution for each event type separately and for the standard case without event types, for the CTAO Southern array. They all correspond to point-like IRFs, in which a direction cut optimizing point-source sensitivity has been applied. Effective area and background rate also show the sum of the four event types. Angular and energy resolution also show the ratio of each event type result with respect to the standard case. All of these are for offset angle between 0 and 1 degree.}
    \label{fig:IRFs_s}
\end{figure*}

Figures \ref{fig:IRFs_n} and \ref{fig:IRFs_s} show the IRFs for each event type in comparison to IRFs without event types for both arrays. The following conclusions can be drawn:
\begin{itemize}
    \item Effective areas are, as expected, lower for each event type individually, but their sum is significantly higher than the effective area without event types. This implies that the event-type analysis is more efficient in terms of information use. Many more signal events are surviving the cuts and participating in this kind of analysis, which justifies the improvement in combined sensitivity seen in Section~\ref{sec:res_sensitivity}.
    \item  Background rates show that most background events tend to be tagged as the bad quality event types. This is also the expected behavior, as the direction reconstruction algorithms are only meant for gamma-ray events.  As a consequence, in addition to the separation by reconstruction quality, our method separates most of the background events prior to the \textit{gammaness} cut optimization. This is also beneficial as the cuts will be more ``fine-tuned'' for each event type.
    \item Angular resolution is where the difference between event types is more remarkable, as event misdirection was the parameter used to define these event types.
    \item Energy resolution also shows a significant difference between event types, even if energy reconstruction quality was never used to define these event types. This highlights the strong correlation between well reconstructed events in direction and in energy.
\end{itemize}

\begin{figure*}
    \centering
    \includegraphics[width=0.8\linewidth]{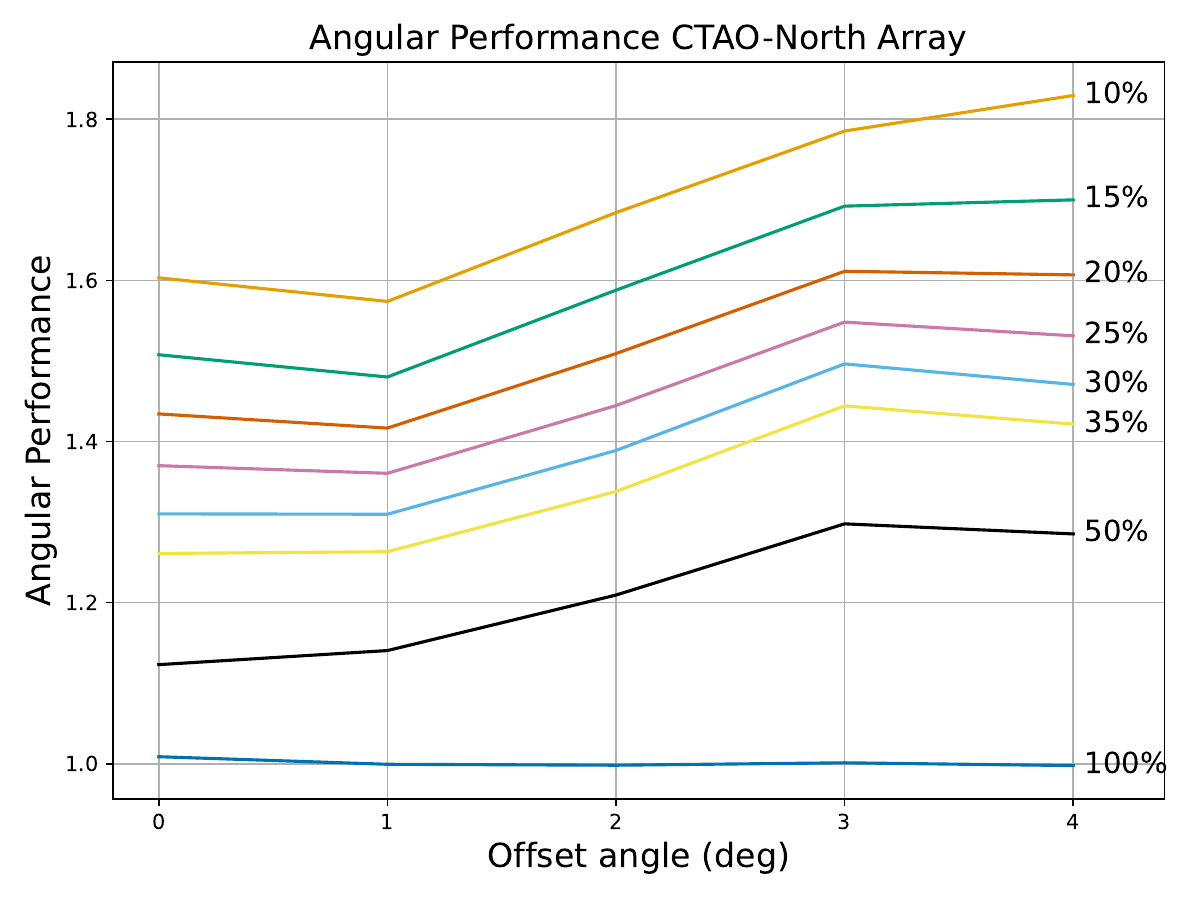}
    \includegraphics[width=0.8\linewidth]{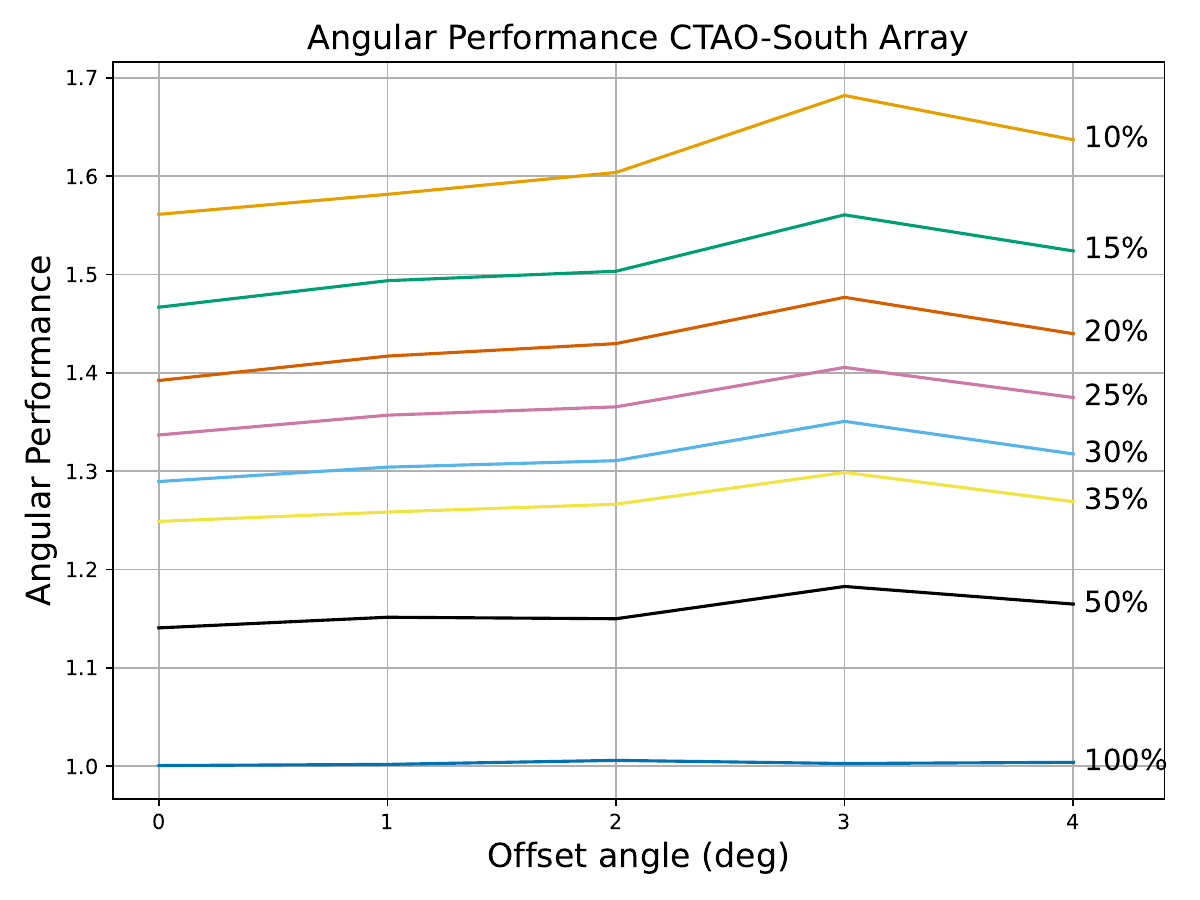}
    \caption{Angular Performances (APs) for the angular resolution of an event type 1 containing the top n\% of gamma-rays ranked according to expected misdirection, using as reference the angular resolution without event types. Results shown for both north and south arrays.}
    \label{fig:APs}
\end{figure*}

Different partitions of event types differ mainly in the angular resolution of each type. Similarly to the PPUTs calculated for the sensitivity, we calculated the APs, as detailed in Section~\ref{sec:hl_sensitivity}, for the angular resolution of the "event type 1" in each case. Figure \ref{fig:APs} shows how the angular resolution changes when selecting the top gamma-ray events, ranked according to their expected misdirection. For instance, the gamma-ray events identified within the top 10\% lead to an improved angular resolution of up to 80\% with respect to the standard analysis. This demonstrates the capability of this method to select events with the best angular resolution, and it also shows there is information that would otherwise be lost and that can be exploited by the event-type-based analysis to improve the resolving power of the CTAO. As shown here, an event-type-based analysis will have an even more pronounced impact over large FOV offset angles, showing its potential to improve the CTAO surveying capabilities.

\subsection{Event-type analysis performance: High-level analysis}
\label{sec:res_high_level}

For the following results, we use as input for \textit{Gammapy} the full-enclosure IRFs of the previously selected event type partition (15\%, 15\%, 30\% and 40\%) to test the real impact of this kind of analysis in typical cases where the angular resolution is the key factor, as explained in Sections~\ref{sec:hl_extension}~and~\ref{sec:hl_separation}.

\subsubsection{Extension detection}
\label{sec:res_extension}

\begin{figure*}
    \centering
    \includegraphics[width=0.49\linewidth]{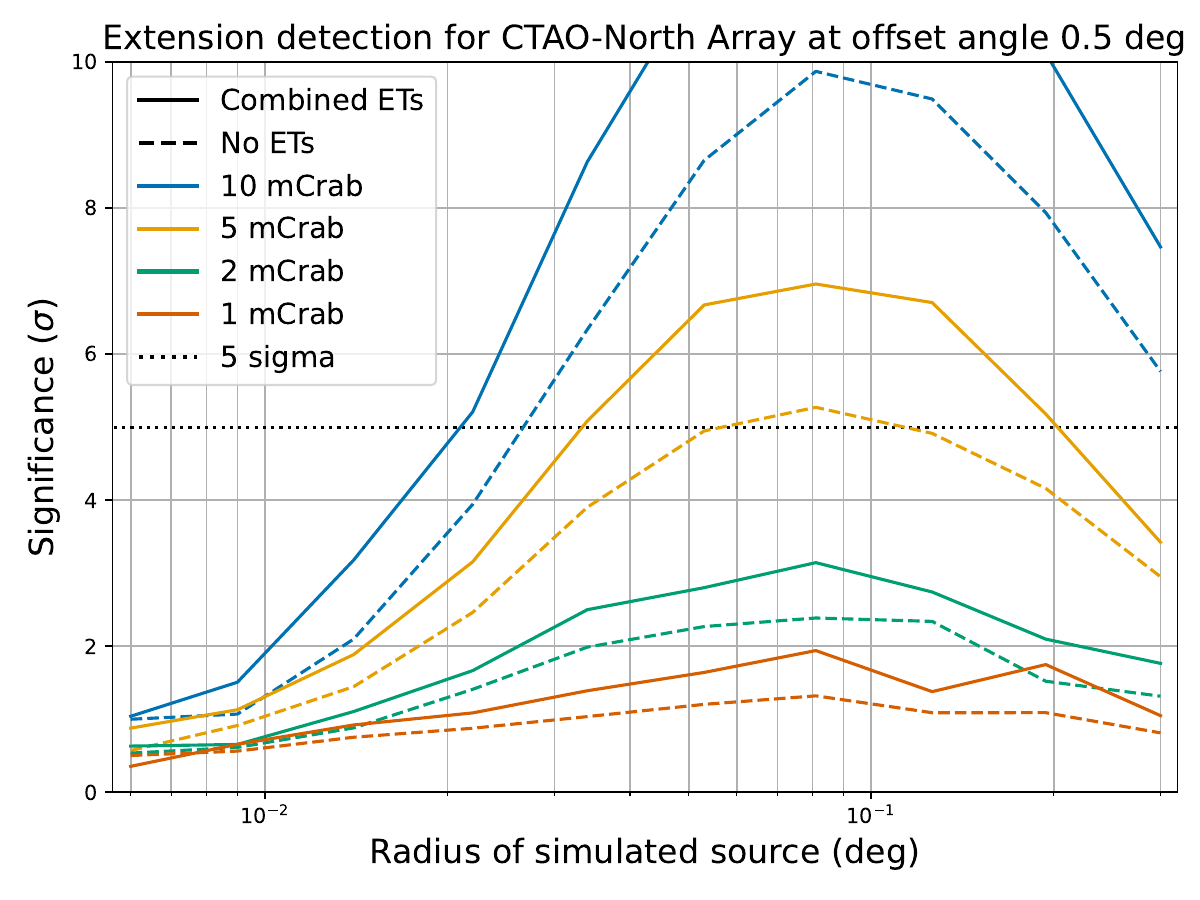}
    \includegraphics[width=0.49\linewidth]{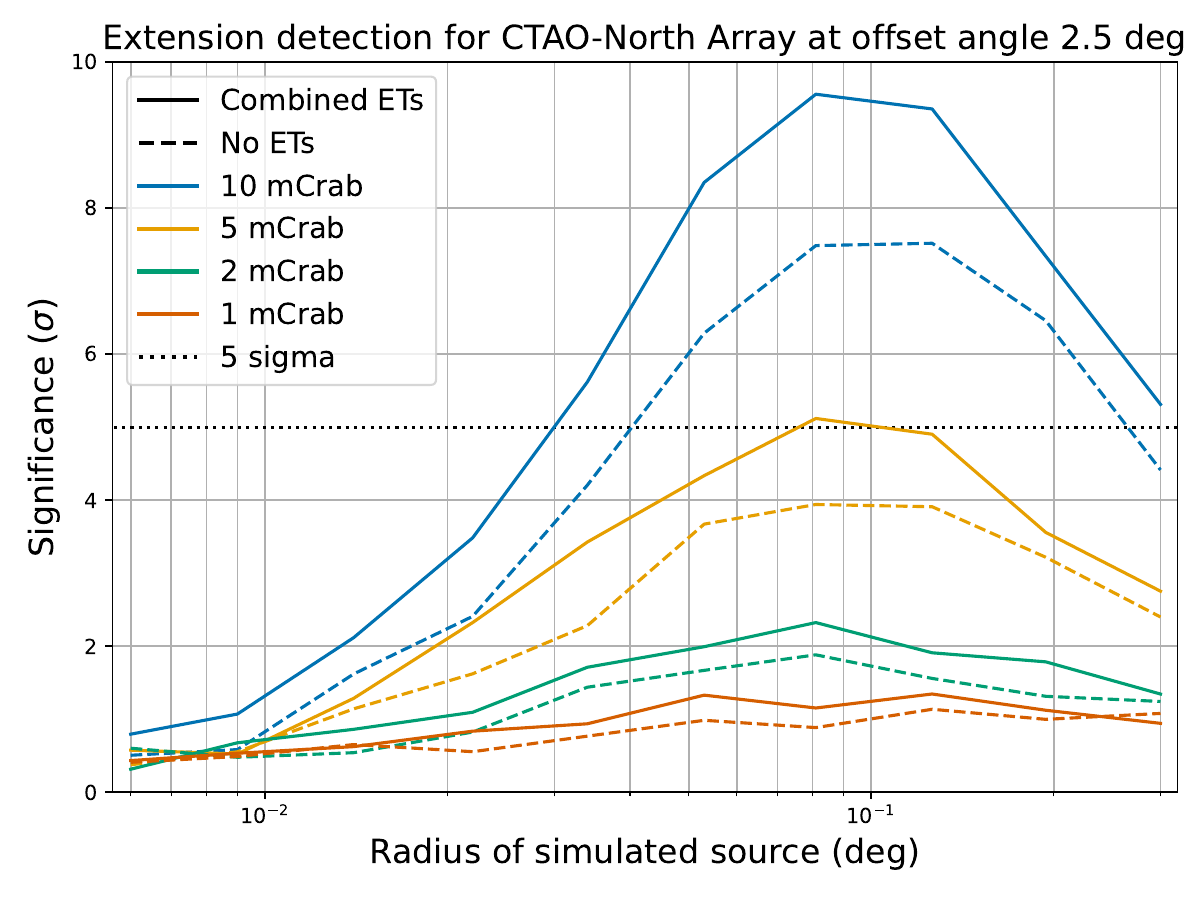}
    \includegraphics[width=0.49\linewidth]{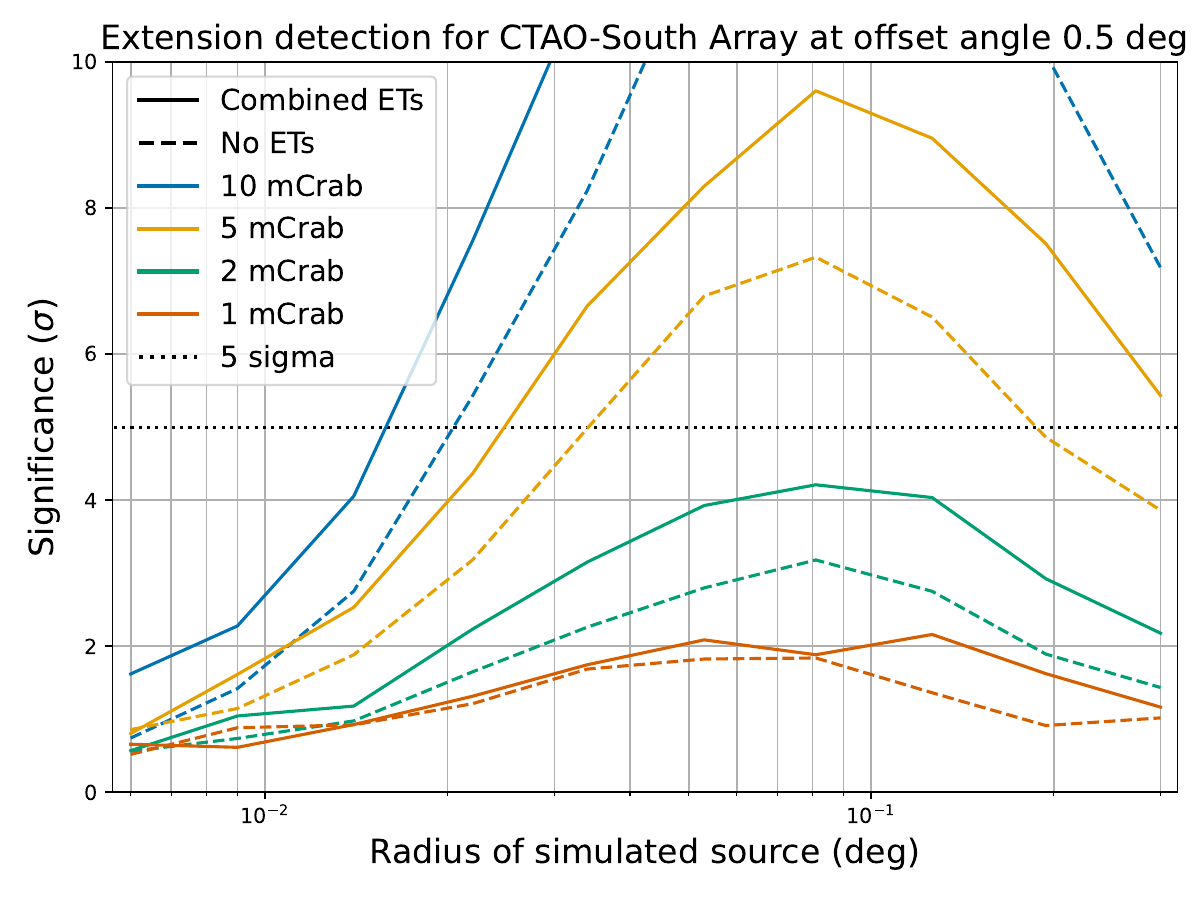}
    \includegraphics[width=0.49\linewidth]{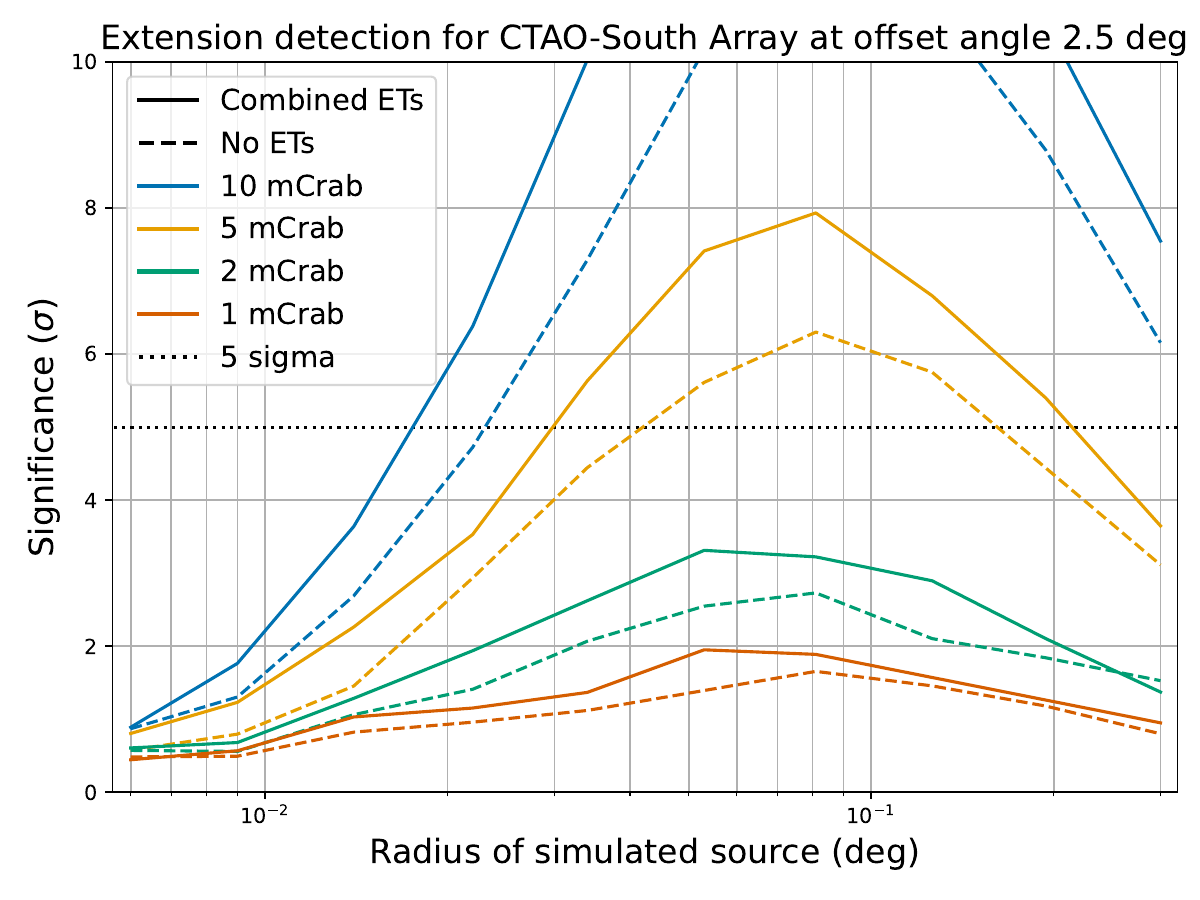}
    \caption{Significance of the detection of the source extension for different radii and intensities of the simulated source. Results for the combination of all four event types (solid lines) and for the standard IRFs without event types (dashed). Each point is the mean significance of 50 repetitions of the simulation and analysis. Repeated at two different offset angles for both arrays.}
    \label{fig:extension}
\end{figure*}

In Figure \ref{fig:extension}, the results of the extension tests for offset angles between 0.5 and 2.5 degrees are shown for both arrays. The intensities simulated in this section were defined as representative examples of source intensities that will be found in the GPS \citep{gps_2024}.

All extension detection probabilities are significantly higher when performing the combined event-type wise analysis, with respect to the standard analysis (with no event types). For sources with an intensity of 5 mCrabs or higher, the extension is detectable for a wider range of radii for all cases where it was already detectable with the standard analysis. For sources of  this intensity at an offset of 2.5 deg in the North Array, the 5-$\sigma$ detection threshold is reached only when using the event-type-based analysis. For the same source amplitude, the detectable extension radius is improved by 25\%. Figure \ref{fig:extension} also shows that the significance for the detection of the extension for sources with the same intensity starts decreasing for radii beyond 0.1 deg. The reason behind this is that in the simulated cases, we kept the intensity (i.e. the amplitude of the spectral model in mCrabs) constant: as the source becomes more extended, the resulting surface brightness gets dimmer, and therefore more difficult to distinguish from the background.

% We can see that for sources with an intensity of 5 mCrabs or higher, the extension is detectable for a wider range of radii of the source in all cases.

\subsubsection{Separation of close sources}
\label{sec:res_separation}

\begin{figure*}
    \centering
    \includegraphics[width=0.49\linewidth]{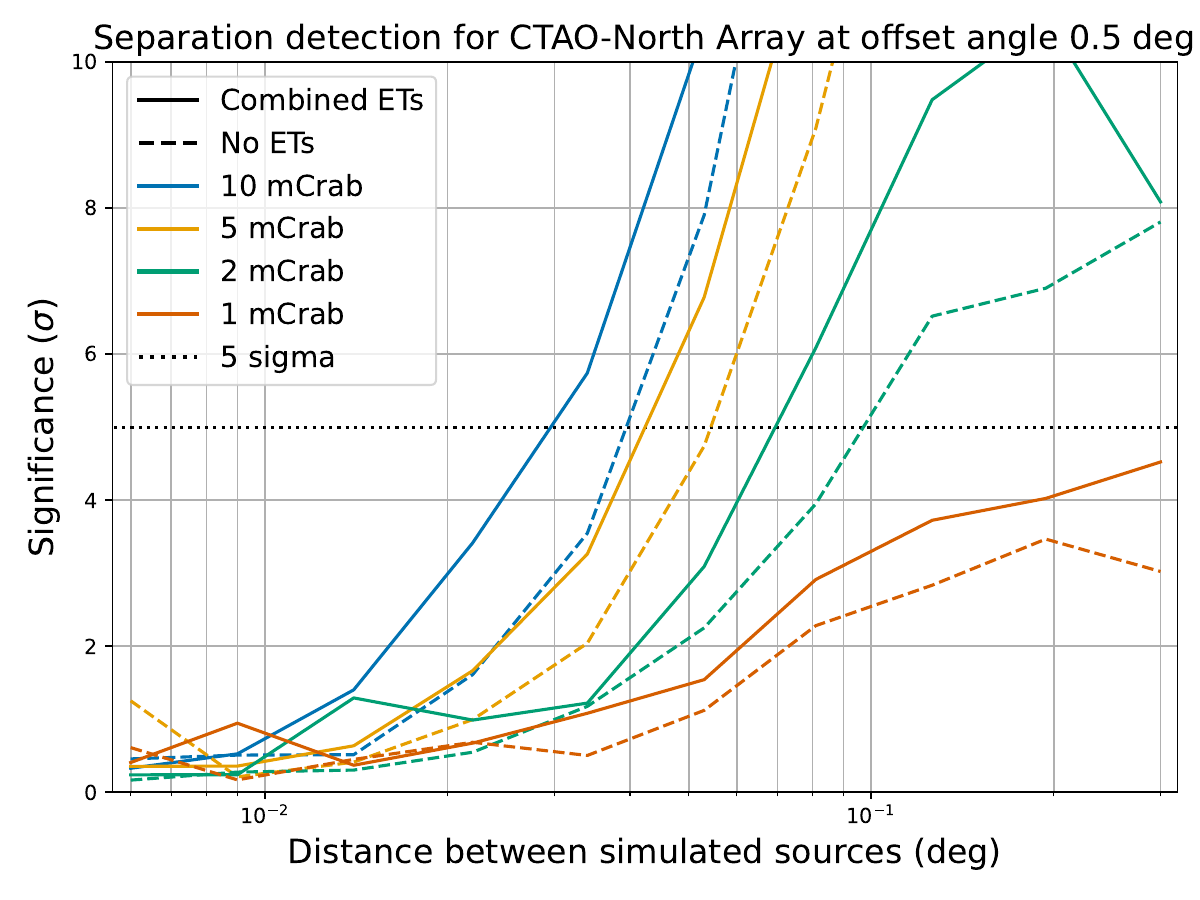}
    \includegraphics[width=0.49\linewidth]{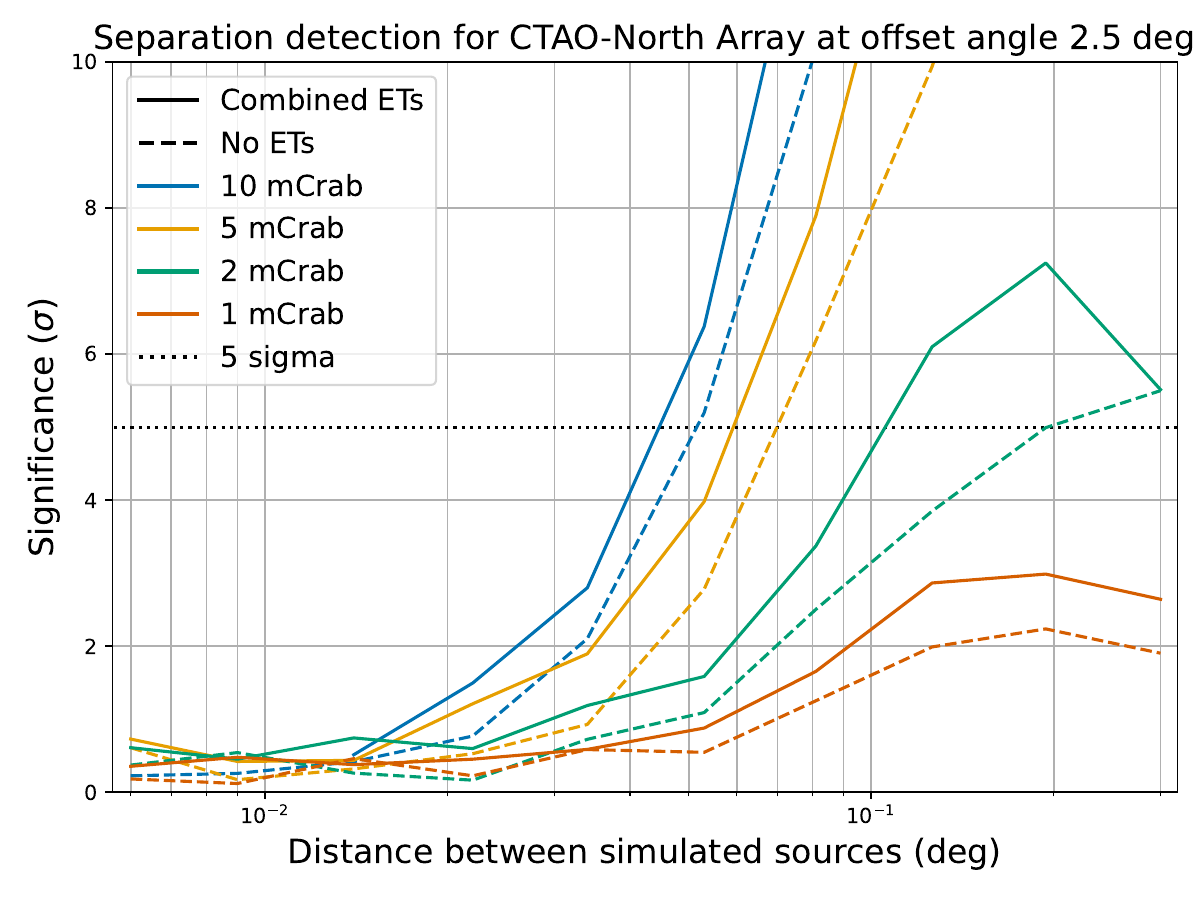}
    \includegraphics[width=0.49\linewidth]{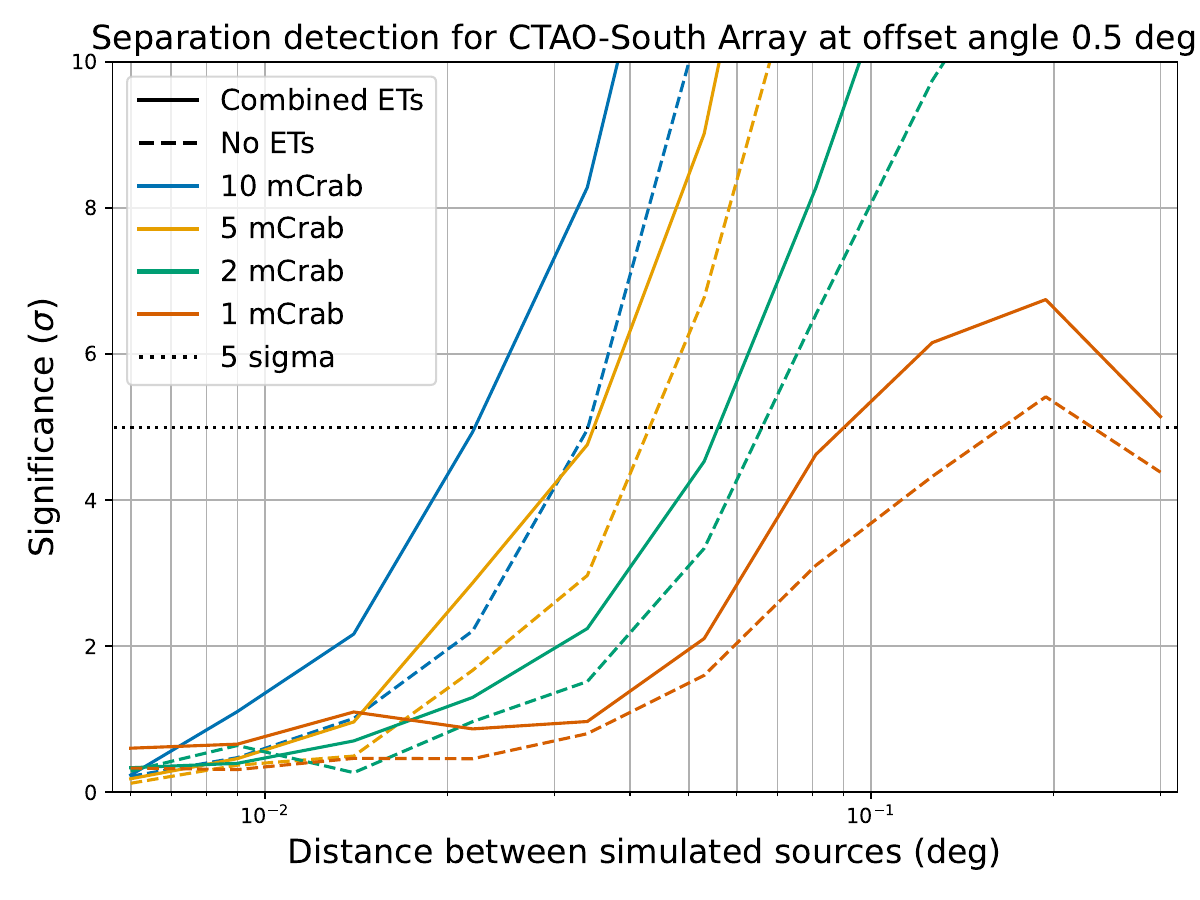}
    \includegraphics[width=0.49\linewidth]{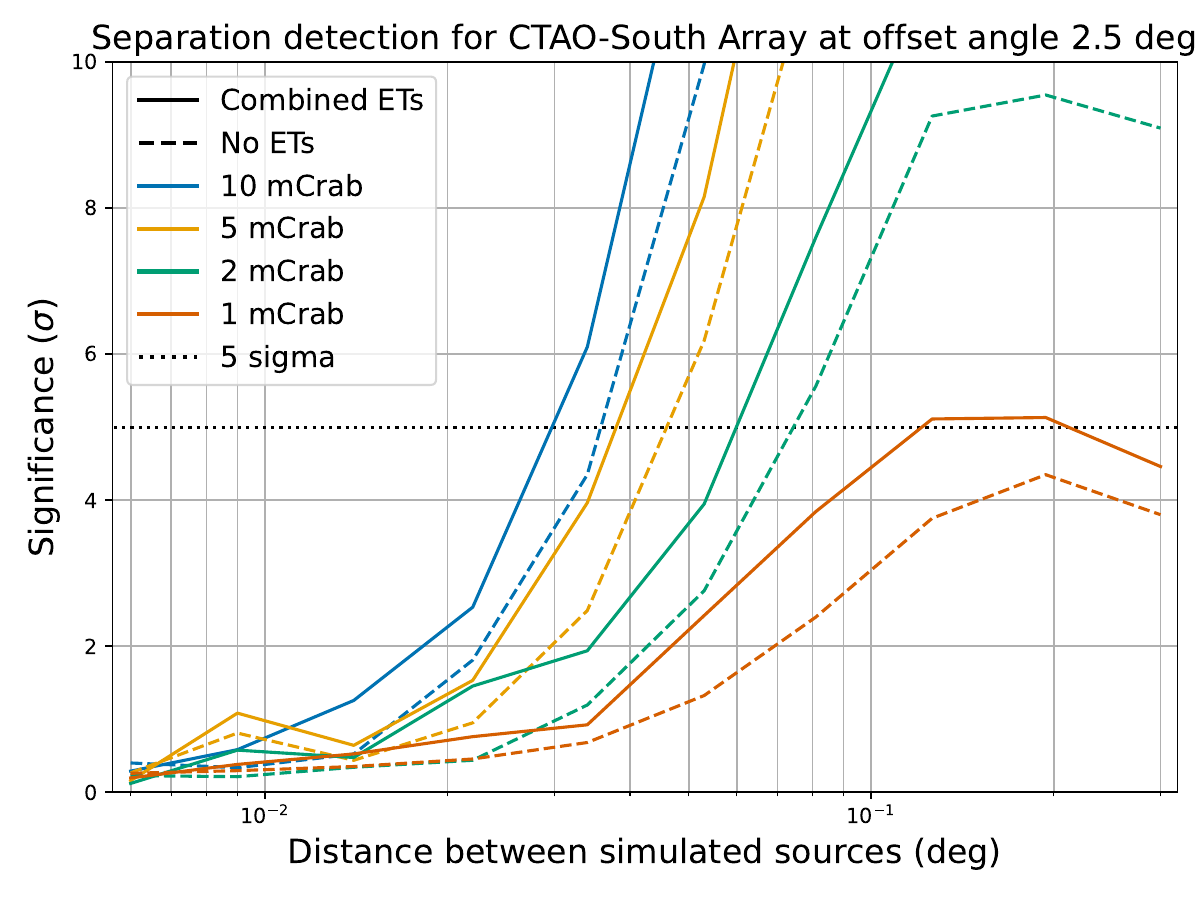}
    \caption{Significance of the detection two separated sources for different separations and intensities of the simulated sources. Results for the combination of all four event types (solid lines) and for the standard IRFs without event types (dashed). Each point is the mean significance of 50 repetitions of the simulation and analysis. Repeated at two different offset angles for both arrays.}
    \label{fig:separation}
\end{figure*}

The results for the separation tests are shown in Figure \ref{fig:separation}. The parameters are the same as in the extension tests, having the distance between two point-like sources instead of the radius of an extended source. 

In this tests, we can also see that all mean significances are higher (up to a factor 1.3) for the combination of all event types than for the analysis without event types.

\section{Conclusions}
\label{sec:conclusions}

The results presented in this work depend both on the low-level analysis carried out (in this case, \textit{EventDisplay}) and the performance of the event-type separation algorithm. In the past, similar approaches have been tested using both \textit{Mars} \citep{2013ICRC_Mars} and \textit{EventDisplay} \citep{maier2017eventdisplay} data samples and different event-type classification algorithms \citep{Hassan_2021}, and in the near future it is expected that the CTAO low-level analysis will be performed with other frameworks (e.g., \textit{ctapipe}). All these studies point out that the main conclusions of this work---namely, the improvements associated with sensitivity, angular resolution, and energy resolution---are present irrespective of the low-level analysis framework chosen. This is expected due to the nature of the IACT technique itself, as certain reconstructed events will have conditions better suited for the reconstruction than others. 

The impact on sensitivity of the number of event types and the particular partition chosen has also been tested. With two event types we obtain an improvement on sensitivity of $\sim$15\% across the tested range of camera offsets and a higher number of event-type separations seem to enhance the performance further. Using 4 event types leads to a boost in sensitivity of $\sim$25\%, but the improvement when increasing farther the number of event-type separations stops being significant beyond this point. These results were were calculated for 50h of exposure time. The scale of the reported improvements in sensitivity becomes lower (in the 5-15\% range) when considering shorter exposure times. Short observations such as transient sources will still benefit from this analysis.

%These improvements, calculated for 50h of exposure time, are lower for shorter exposure times, but still significant, meaning that short observations like transient sources can benefit from this analysis, but not as strongly as longer observations.

It is important to note that all the results presented in this paper are obtained using exclusively MC simulations. On one hand, the reported benefits could diminish when applying this methodology to real data, which may present some significant differences with respect to simulations. The improvement of gamma/hadron separation is especially sensitive to these differences and could be affected. On the other hand, the CTAO has been designed with strict requirements on systematic uncertainties, and therefore data to MC differences are expected to be under control. The natural next step of this work will be to apply the analysis described here to real data from the CTAO telescopes. This can be accomplished using data from the Large-Sized Telescope prototype (LST-1) of the CTAO located in the Northern site at the Roque de los Muchachos Observatory \citep{Morcuende_2023}.

We see a $\sim$25\% improvement in sensitivity, likely due to the better cut optimization performed on events with clearly different associated quality and a boost between 25 to 50\% in resolving power. As the best event type shows a boost in energy resolution between 5 to 20\%, we also expect an improved sensitivity for science cases limited by this parameter, such as sharp spectral cutoffs or line detection. This improvement in energy resolution shows how highly correlated angular and energy reconstruction quality are in the IACT technique. In fact, this correlation presents a problem for likelihood-ratio-test-based analyses (e.g. \textit{Gammapy}, \textit{ctools}), as accounting for it properly would require higher dimensional IRFs including this correlation, making the analysis computationally challenging. An event-type-based analysis, in addition to improving high-level analysis performance, it is also expected to mitigate the effect of this correlation. The evaluation of this and other systematic effects are considered out of scope for this work, and will be explored once real data is available.

If the improvements seen at the MC level are proven to be accurate using telescope prototypes data, the event-type-based analysis will have a very significant impact on most of the scientific objectives of the CTAO. A fraction of these scientific objectives (e.g., IGMF measurements, GPS source separation, DM line searches) are limited by the resolution of the instruments, and therefore we expect that an event-type-based analysis will have an even greater effect on these objectives. This improvement may be large enough to justify a re-evaluation of current estimations.

%In addition to the boost on extended sources detection and surveys, the improvement of angular resolution especially at high offset distances will be highly valuable for observational targets with a large position error.
In addition to the boost on source separation and extension determination, the very-significant improvement of angular resolution at high offset distances will be highly valuable for both surveying programs as well as discovering targets with large positional errors (e.g., rapid follow-up observations of gamma-ray bursts, astrophysical neutrinos, or gravitational-wave events) \citep{cta_book}.

Better ways of integrating an event-type-based analysis in the low-level analysis of the future CTAO need to be explored. The implementation proposed in this paper requires a dedicated machine learning train/test stage, as the algorithm is tailored exclusively to characterize events according to their expected angular reconstruction performance. Other reconstruction methods \citep[e.g.][]{schwefer_2024} have these capabilities built-in, which means no additional MC statistics would be required to define event-type partitions.

%%%%%%%%%%%%%%%%%%%%%%%%%%%%%%%%%%%%%%%%%%%%%%%%%%
\section*{Acknowledgements}

This work is partially supported by the Spanish Research State Agency (AEI) through the grants PID2019-104114RB-C31 and PID2022-138172NB-C41 and the Ministry of Science through the budget lines 28.06.000X.411.01 and 28.06.000X.711.04 of PGE 2023, 2024 and 2025. Funded/Co-funded by the European Union (ERC, MicroStars, 101076533). Views and opinions expressed are however those of the author(s) only and do not necessarily reflect those of the European Union or the European Research Council. Neither the European Union nor the granting authority can be held responsible for them. This research used computing and storage resources provided by the Port d’Informació Científica (PIC) data center. GM acknowledges support from DESY (Germany), a member of the Helmholtz Association HGF.

%%%%%%%%%%%%%%%%%%%%%%%%%%%%%%%%%%%%%%%%%%%%%%%%%%
\section*{Author Contributions}
This publication is the result of a collaborative effort of all co-authors. Early design of the project was born within the CTAO Consortium, and first preliminary results were led by TH, OG, GM, ML, M. Peresano, I. Vovk, members of the Analysis and Simulations Working Group. JB and SGS led the drafting of the paper, while all co-authors contributed to the final version. GM produced custom DL2 Monte Carlo data including lower-level features needed in this analysis. OG and TH developed the first version of the machine learning classifiers to determine event types. JB, SGS and OG optimized machine learning methods to maximize classification accuracy. JB and ML implemented the modifications within the pyirf library needed to compute event-type-wise IRFs. JB and AS implemented the high-level analysis performed with gammapy, to evaluate the impact of event types over several performance parameters.

%%%%%%%%%%%%%%%%%%%%%%%%%%%%%%%%%%%%%%%%%%%%%%%%%%
\section*{Data Availability}

Prod-5 IRFs are public, and can be found in \cite{cta_irf_2021}, while event-type-wise IRFs as well as lower-level data products (e.g. DL2 event lists) may be provided upon request. The event-type reconstruction software can be found in \cite{iact_event_types}.

\appendix

\section{Variables used to train the model}
\label{app:train_features}

The input features used to train the neural networks are listed and described in this appendix. Definitions follow the \textit{EventDisplay} framework, with most variables based on second-moment image parameterization \citep{1985ICRC....3..445H}. In some cases, derived quantities are used.

\begin{itemize}
  \item \texttt{log\_EmissionHeight}: Logarithm of the height of the Cherenkov photon emission maximum.
  \item \texttt{array\_distance}: Distance of reconstructed core position to observatory center
  \item \texttt{camera\_offset}: Angular distance of reconstructed source position from camera center
  \item \texttt{log\_reco\_diff}: Logarithm of difference between two direction reconstruction methods (based on dispBDTs and based on the intersection reconstruction methods)
  \item \texttt{log\_NTels\_reco}: Logarithm of the number of telescopes used in the event reconstruction.
  \item \texttt{log\_SizeSecondMax}: Logarithm of the second-largest image size in the event.
  \item \texttt{img2\_ang}: Average angle between image length axes.
  \item \texttt{NTrig}: Number of telescopes participating in the array trigger.
  \item \texttt{MSCL}: Mean reduced scaled image length \citep{2006APh....25..380K}.
  \item \texttt{log\_EmissionHeightChi2}: Logarithm of squared differences between reconstruction height of Cherenkov photon emission maximum using pairs of telescopes only.
  \item \texttt{meanPedvar\_Image}: Mean pedestal variation per telescope.
  \item \texttt{log\_DispDiff}: Logarithm of differences in the reconstruction arrival direction between the results of the single-telescope analysis.
  \item \texttt{sum\_loss}: Sum of the \texttt{loss} parameter. Loss of the fraction of image size in camera edge pixels.
  \item \texttt{MSCW}: Mean reduced scaled image width \citep{2006APh....25..380K}.
  \item \texttt{log\_dESabs}: Logarithm of absolute error in reconstruction energies obtained from single-telescope analysis.
  \item \texttt{log\_reco\_energy}: Logarithm of reconstructed energy (TeV).
  \item \texttt{MSWOL}: Ratio of \texttt{MSCW}/\texttt{MSCL}.
\end{itemize}

For the following parameters, their average (av\_), median (me\_), and standard deviation (std\_) have been used as training features:

\begin{itemize}
  \item \texttt{size}: Total image size per telescope. The logarithm of  \texttt{av\_size},  \texttt{me\_size} and  \texttt{std\_size} is used for the training.
  \item \texttt{cross}: Distance from image centroid to position reconstructed via intersection method.
  \item \texttt{R}:  Distance from each telescope to the reconstructed shower core.
  \item \texttt{dist}: Distance of the image centroid to the camera center.
  \item \texttt{tgrad\_x}: Timing gradient along the image’s major (length) axis.
  \item \texttt{asym}: Measure of shower skew (asymmetry).
  \item \texttt{loss}: Fraction of image size in edge pixels.
  \item \texttt{ES}: Reconstructed energy per telescope.
  \item \texttt{fui}: Fraction of image/border pixel under estimate image ellipse.
\end{itemize}

The following figure (\ref{fig:importance}) shows the importance of every variable used to train the \texttt{MLP\_tanh} model for each energy bin.

\begin{figure*}
\resizebox{\hsize}{!}
        {\includegraphics{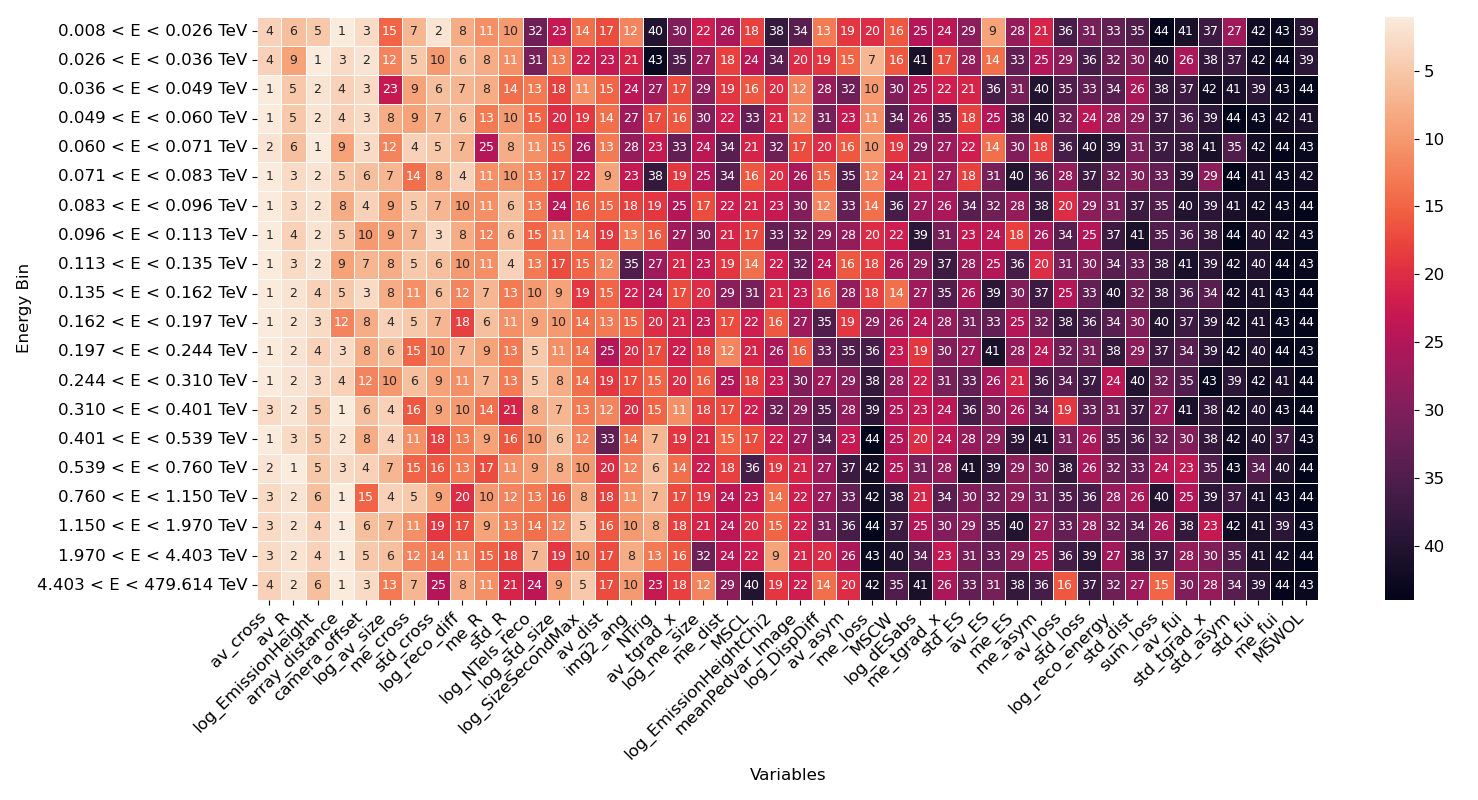}} 
  \caption{Grid showing the ranking of importance calculated using Garson's algorithm. Annotations in the grid represent the rank of that particular variable and energy bin, with the x axis ordered by the overall importance. The most important variables (top ranks) are shown in light orange colors, those with lower importance in purple/black.}
     \label{fig:importance}
\end{figure*}   




\bibliographystyle{elsarticle-num-names}
\bibliography{bibliography}

\end{document}